\documentclass[aps, prl, superscriptaddress, twocolumn, nofootinbib,
notitlepage,  longbibliography, 
floatfix]{revtex4-2}

\usepackage[utf8]{inputenc}
\usepackage{natbib}
\usepackage{graphicx}
\usepackage{mathtools}
\usepackage{amsfonts} 
\usepackage{amsmath} 
\usepackage{amssymb}
\usepackage{amsthm}
\usepackage{bm}
\usepackage{braket}
\usepackage{color} 
\usepackage{bbm}
\usepackage{hyperref}
\usepackage{subcaption}
\usepackage{multirow}
\usepackage{import}
\usepackage{microtype}
\usepackage{relsize}
\usepackage{framed}
\usepackage[dvipsnames]{xcolor}
\usepackage[percent]{overpic}
\usepackage{tikz}
\usepackage{xparse}

\ExplSyntaxOn
\NewDocumentCommand{\colvector}{m}
{
  \begin{pmatrix}
    \clist_use:nn {#1} { \\ }
  \end{pmatrix}
}
\ExplSyntaxOff

\newcommand{\Z}{\mathbb{Z}}

\newtheorem*{corollary*}{Corollary}

\theoremstyle{definition}

\usepackage{newtxtext}
\usepackage{CJK}

\usepackage{comment}
\usepackage{ragged2e}
\usepackage{framed}
\usepackage{physics} 
\usepackage{dsfont}
\usepackage[normalem]{ulem}
\usepackage{placeins}
\usepackage{subcaption}
\definecolor{darkblue}{rgb}{0.,0.,0.4}

\begin{document}

\author{Yasamin Panahi}
\email{ypanahi@pitp.ca}
\affiliation{Perimeter Institute for Theoretical Physics, Waterloo, Ontario N2L 2Y5, Canada}
\affiliation{Department of Physics and Astronomy, University of Waterloo, Waterloo, Ontario N2L 3G1, Canada}

\author{Subhayan Sahu}
\email{ssahu@pitp.ca}
\affiliation{Perimeter Institute for Theoretical Physics, Waterloo, Ontario N2L 2Y5, Canada}

\author{Naren Manjunath}
\email{nmanjunath@pitp.ca}
\affiliation{Perimeter Institute for Theoretical Physics, Waterloo, Ontario N2L 2Y5, Canada}

\author{Chong Wang (王翀)}
\email{cwang4@pitp.ca}
\affiliation{Perimeter Institute for Theoretical Physics, Waterloo, Ontario N2L 2Y5, Canada}

\title{Quantum criticality at strong randomness: a lesson from anomaly}

\begin{abstract}
 Quantum criticality in the presence of strong quenched randomness remains a challenging topic in modern condensed matter theory. We show that the topology and anomaly associated with \textit{average symmetry} can be used to predict certain nontrivial universal properties. Our focus is on systems subject to average Lieb--Schultz--Mattis constraints, where lattice translation symmetry is preserved only on average, while on-site symmetries remain exact. We argue that in the absence of spontaneous symmetry breaking and intrinsic topological order, the system must exhibit critical correlations of local operators in two distinct ways: (i) for some operator $O_e$ charged under exact symmetries, the \textit{first absolute moment} correlation $\overline{|\langle O_e(x)O^{\dagger}_e(y)\rangle|}$ decays slowly; and (ii) for some operator $O_a$ charged under average symmetries, the \textit{first-moment} correlation $\overline{\langle O_a(x)O^{\dagger}_a(y)\rangle}$ decays slowly. We verify these predictions in a few examples: the random-singlet Heisenberg spin chain in one dimension, and the disordered free-fermion critical states in symmetry class BDI in one and two dimensions. Surprisingly, even for these well-studied systems, our anomaly-based argument reveals critical correlations overlooked in previous literature. We also discuss the experimental feasibility of measuring these critical correlations.
 
\end{abstract}

\begin{CJK*}{UTF8}{bkai}
\maketitle
\end{CJK*}

%\tableofcontents

%\section{Introduction}

Quantum criticality in the presence of strong quenched randomness is ubiquitous in modern condensed matter physics, from superconductor-insulator transition \cite{FisherWeichmanGrinsteinFisher1989,Sondhi1997QPT}, to quantum Hall effects \cite{Huckestein1995QHE,AndersonTransitionRMP} and disordered quantum spin systems \cite{MaDasguptaHu1979,PhysRevB.50.3799,IgloiMonthus2005}. Despite its long history, the subject of \textit{disordered quantum criticality} has remained extremely challenging for theoretical understanding. The difficulty stems from the lack of theoretical tools that are both non-perturbative and broadly applicable. 

The notion of symmetry anomaly (more precisely, 't Hooft anomaly) \cite{tHooft1979,else_classifying_2014}
has become central in our non-perturbative understanding of both condensed matter physics and quantum field theory. In condensed matter systems, the most common manifestation of symmetry anomaly is the Lieb-Schultz-Mattis (LSM) constraint \cite{LiebSchultzMattis1961,Oshikawa1999LSM,Hastings2004LSM} and its many generalizations \cite{tasaki2022LSMreview,Cheng2016LSM,Po2017,Else2020}. LSM-type constraints require the ground states to be nontrivial (namely long-range entangled \cite{ChenGuWenLRE}) for any lattice systems with translation symmetry and ``fractional'' symmetry quantum numbers (\textit{e.g.} spin-$1/2$ under $SO(3)$) per unit cell~\cite{GioiaWangMomentum,LYZZ2025}. Furthermore, they also heavily constrain what type of nontrivial low energy theory can emerge, since the anomaly of the low energy theory (IR) must match that of the microscopic lattice system (UV). A familiar example of this type is the Luttinger theorem, which demands the Fermi surface volume (an IR notion) to match with the microscopic charge density (a UV notion)~\cite{Oshikawa2000Luttinger,ElseThorngrenSenthil2020}. Examples of such LSM anomaly-matching with conformal field theory (CFT) in the IR have also been discussed extensively in both one and two dimensions \cite{Cho2017LSM,Jian2018lsm,Metlitski2018,Zou2021,CalveraWang2021,Cheng2023LSM,Lanzetta_2023,Seiberg2025LSM,ZhangSongSenthil2025}.

Although 't Hooft anomalies have played a central role in many areas of modern condensed matter and high-energy physics, they have seen relatively limited use in the study of disordered quantum criticality. An isolated example of an LSM-type constraint in a one-dimensional spin chain with exact $SO(3)$ spin rotation symmetry and average translation symmetry was discussed in Ref.~\cite{KimchiNahumSenthil}. More generally, the notion of \textit{average anomaly} was introduced in Ref.~\cite{MaWang2023} and systematically classified in Ref.~\cite{MZBCW2025}. An average anomaly typically involves both an \textit{exact symmetry} -- one that holds for every disorder realization -- and an \textit{average symmetry} -- one that is preserved only upon averaging over the disorder ensemble. It was argued in Ref.~\cite{MaWang2023} that systems with such average anomaly must have long-range entangled ground states.

Despite its early successes, the potential of the anomaly-based approach remains far from fully realized. In particular, can quantum anomalies be used to gain a deeper understanding of the universal properties of certain disordered quantum critical states? A closely related issue is that, although the notion of long-range entanglement -- a consequence of anomaly -- is conceptually appealing, it is not a property that can be directly and efficiently measured in experiments. It is therefore desirable to formulate predictions directly in terms of measurable quantities, such as correlation functions.

In this letter we uncover how anomalies constrain correlation functions for disordered quantum critical states. Here, by a disordered quantum critical state we mean a nontrivial gapless IR state arising when the anomaly is not realized by spontaneous symmetry breaking or intrinsic topological order. As an example, we consider the \textit{average LSM} anomaly in general $d$ space dimension, where lattice translation is an average symmetry, while some on-site symmetry (like $SO(3)$, time-reversal or fermion parity) remains an exact symmetry and is represented projectively (\textit{e.g.} a spin-$1/2$ for $SO(3)$, Kramers' doublet for time-reversal or unpaired Majorana for fermion parity) in the Hilbert space of a unit cell; this property defines the average LSM anomaly. We choose to focus on LSM anomalies as they are the most commonly encountered anomalies in condensed matter systems. Our discussion, however, holds more generally for anomalies involving both exact and average symmetry. The central result of this work is the following:

% \begin{framed}
% {\textit{Power-law rule}: if a system has anomalous exact and average symmetry that are not spontaneously broken in the ground state, then for some operator $O_e$ charged under exact symmetries, the \textit{first absolute moment} correlation $\overline{|\langle O_e(x)O^{\dagger}_e(y)\rangle|}$ (known as the Edwards--Anderson or EA correlator) decays critically; for some operator $O_a$ charged under average symmetries, the \textit{first-moment} correlation $\overline{\langle O_a(x)O^{\dagger}_a(y)\rangle}$ decays critically. Here ``critical" decay typically means power-law $\sim 1/|x-y|^{2\Delta}$.}
% \end{framed}
{\itshape \textbf{Power-law rule}: if a system has anomalous exact and average symmetry that are not spontaneously broken in the ground state, and if the anomaly is not saturated by intrinsic topological order, then there exists an operator $O_e$ charged under exact symmetries whose \textit{first absolute moment} correlation $\overline{|\langle O_e(x)O^{\dagger}_e(y)\rangle|}$ (known as the Edwards--Anderson or EA correlator) decays critically; and there exists an operator $O_a$ charged under average symmetries whose \textit{first-moment} correlation $\overline{\langle O_a(x)O^{\dagger}_a(y)\rangle}$ decays critically. Here ``critical" decay typically means power-law $\sim 1/|x-y|^{2\Delta}$.}\\

For typical critical phases studied in this work, the symmetry-charged operators $O$ are local. There can be nontrivial exceptions: for example, if the system develops nontrivial topological orders, the charge operator $O$ may act on large strings or higher dimensional sub-manifolds and the correlation function will be non-decaying.

We emphasize at the outset that we do not have a rigorous proof of our power-law rule -- such results are difficult to establish even in clean systems, especially in general dimensions, although they are widely expected to hold. Instead, we present plausibility arguments and illustrate them through several nontrivial examples: the random-singlet phase in $1d$ spin chains~\cite{MaDasguptaHu1979, DasguptaMa1980, PhysRevB.50.3799}, and the disordered free-fermion critical states in symmetry class BDI, in both one and two dimensions. The one dimensional state is related to the random transverse-field Ising model~\cite{Fisher1995Ising}, and the two dimensional state, known as the Gade phase, has been extensively studied~\cite{Gade1991, Gade1993, Ludwig1994,Mudry2003, Hafner2014, Sanyal2016, Konig2012, Karcher2023, Nayak2024, Das2025Z2FluxDisorder}. 
Surprisingly, even for these relatively well-studied examples, our results predict nontrivial correlation functions that have been overlooked in the prior literature. Specifically, we show that in both examples, the first-moment dimer-dimer correlation function (see Eq.~\eqref{eq:Dfm-def}) exhibits a nontrivial power-law decay.

\section*{Anomaly and Correlation}
Here we give a plausibility argument for our main claims, the ``power-law rule'', about how anomaly constrains correlation functions. Let us first recall how such constraints operate in clean systems. In a clean system with global symmetry $G$, a nontrivial 't Hooft anomaly guarantees that the system cannot realize a trivially gapped ground state. Equivalently, assuming a sensible notion of renormalization-group (RG) flow, a theory with a nontrivial anomaly cannot flow to a trivial infrared (IR) effective theory. For concreteness, recall two familiar examples of nontrivial 't Hooft anomalies in condensed matter physics: (i) the boundary of a topological insulator (in two or three dimensions), where gapless edge modes are guaranteed as long as charge-conservation $U(1)$ and time-reversal $\Z_2^T$ symmetries remain unbroken; (ii) spin systems subject to LSM constraints (half-integer $SO(3)$ spin per unit cell) in arbitrary spatial dimension $d$, where the ground state cannot be trivial as long as spin-rotation $SO(3)$ and lattice translation symmetry $\Z^d$ remain unbroken. 

In fact, an anomaly implies much more than merely a nontrivial IR theory. As an invariant under RG flow, a nontrivial anomaly must also be reflected in the effective IR description. This immediately implies that the symmetry involved in the anomaly must act nontrivially in the IR, in the sense that certain operators $\{O\}$ carry charges under the anomalous symmetry and exhibit nontrivial correlation functions in the IR. Typically, $\{O\}$  is a set of local operators, and a ``nontrivial'' correlation function means a power-law decay 
\begin{equation}
\label{eq:powerlaw}
    \langle O(x)O^{\dagger}(y)\rangle\sim \frac{1}{|x-y|^{2\Delta_O}}.
\end{equation}
The rationale behind the above ``power-law rule'' is that if, contrary to this rule, all charged operators had uninteresting correlation functions -- such as exponential decay -- then the symmetry would effectively act trivially in the IR theory, in contradiction with the RG invariance of the 't Hooft anomaly. We can again recall some familiar examples: for the gapless Dirac fermions on the boundary of a topological insulator, $\{O\}$ includes the Dirac mass (odd under time-reversal) and the Cooper pair (charged under $U(1)$); for the spin-$1/2$ antiferromagnetic Heisenberg chain, $\{O\}$ includes the spin operator $\vec{S}_i$ ($i$ labeling the lattice site), which is charged under $SO(3)$, and the staggered dimer operator $(-1)^i\vec{S}_i\cdot\vec{S}_{i+1}$, which flips sign under translation by one unit cell $T_x$.\footnote{Interestingly, in the Heisenberg-chain example, all operators with power-law correlations transform trivially under translation by two unit cells $T^2_x$. In other words, all nontrivial IR operators carry lattice momentum either $0$ or $\pi$. This is fully compatible with the anomaly: the $\Z_2$ nature of the LSM anomaly implies that symmetry operations like $T_x^2$ are not involved in the anomaly and are therefore allowed to act trivially in the IR theory. Of course, this does not forbid $T_x^2$ from acting nontrivially on some other IR theory of the system with a very different lattice Hamiltonian.}

We note that not all anomalous theories satisfy the above power-law rule. Notable exceptions include spontaneous-symmetry-breaking (SSB) phases and gapped topological orders. In such cases, the anomalous symmetry can act in more subtle ways, not necessarily through local operators with power-law correlations. For example, there may exist nontrivial gapped excitations -- such as domain walls in SSB phases or anyons in topological orders -- that transform nontrivially under the anomalous symmetry. In this work we restrict attention to systems that do not enter SSB or topologically ordered phases -- phases that are, in any case, much simpler than quantum criticality -- and are therefore expected to obey the power-law rule. We emphasize again that, even for clean systems, the power-law rule has no rigorous proof in general, even though it is widely expected to be true.

We now consider disordered systems with total symmetry $\mathcal{G}=K\times G$, where $K$ is an exact symmetry and $G$ is only an average symmetry (more general forms of symmetry can be treated similarly~\cite{MZBCW2025}). We shall consider examples where $K$ and $G$ have a mixed 't Hooft anomaly, namely the symmetry is anomalous if and only if both $K$ and $G$ are unbroken. A physically relevant example, which will be the main focus of this work, is LSM-type of anomalies with an average lattice translation symmetry. It was shown~\cite{MaWang2023} that as long as the on-site symmetry ($SO(3)$, time-reversal or fermion parity) remains exact and is represented projectively (spin-$1/2$, Kramers doublet, unpaired Majorana fermion) in each unit cell, the LSM anomaly remains nontrivial in this setting. If the on-site symmetry is broken, even only to an average symmetry by a random magnetic field, the anomaly becomes trivial, since the ground state in each disorder realization can now be a trivial product state (with the spins simply aligning with the local random field). Purely internal average symmetries can also realize mixed average anomalies, although the allowed disorder is more constrained. For example, in decorated-domain-wall constructions, defects of the average symmetry carry projective representations of the exact symmetry $K$; statistically $G$-symmetric, $K$-preserving disorder can randomize the defect configuration while preserving these projective $K$ quantum numbers, leaving the anomaly to be matched by SSB, intrinsic topological order, or a gapless IR state.

We  now discuss the power-law rule  for systems with nontrivial average anomalies. A natural question immediately arises: what type of correlation function should we consider? Candidates include the first-moment correlation $\overline{\langle O O^{\dagger}\rangle}$ and Edwards--Anderson-type correlation $\overline{|\langle O O^{\dagger}\rangle|}$, where $\langle...\rangle$ denotes the quantum expectation value in a single disorder realization, and $\overline{\cdot \cdot\cdot}$ denotes averaging over the disorder ensemble. Here we claim that for operators charged under the exact symmetry, we should consider the Edwards--Anderson (EA) correlator; while for operators charged under the average symmetry, it suffices to consider the first-moment correlator. In the Supplementary Material~\cite{SM_RandomLesson}, we provide a derivation of this result using the replica trick. Here we shall provide a physically intuitive argument.

The key observation is that the power-law rule, Eq.~\eqref{eq:powerlaw}, indicates that the symmetry is on the verge of spontaneous breaking. Recall that a power-law decay of a charged operator is often referred to as ``quasi-long-range order'' in Berezinskii--Kosterlitz--Thouless (BKT) physics, while true symmetry-breaking order corresponds to the limit $\Delta_O=0$. Therefore we can infer which correlators should satisfy a power-law rule based on which correlator  detects the limiting case of symmetry-breaking order. 

Let us first examine the average symmetry $G$. What does it mean to have $G$ spontaneously broken? Consider a local operator $O_a$ that is charged under the average symmetry $G$, and an explicit breaking of $G$ would mean $\overline{\langle O_a(x)\rangle}\neq0$. Consequently, a spontaneous breaking of $G$ can be detected through the two-point correlation function 
\begin{equation}
\label{eq:1stcorrelator}
    C_a(|x-y|) = \overline{\langle O_a(x) O^\dagger_a(y)\rangle},
\end{equation}
which approaches a nonzero constant as $|x-y|\to\infty$ in the SSB phase. In our discussion of anomaly constraints, we assume that $G$ is not spontaneously broken, so $C_a(|x-y|)$ does not saturate to a nonzero constant and instead decays to zero at large distances.
Therefore, we come to the expectation stated as the power-law rule, that there exists a $G$-charged operator $O_a$ for which $C_a(|x-y|)$ decays as a power-law. The power-law decay of the first-moment correlator can be viewed as a quasi-long-range order where the average symmetry $G$ is on the verge of spontaneous breaking. Notably, being a first-moment quantity makes the correlation function Eq.~\eqref{eq:1stcorrelator} feasible for experimental observation -- a point we shall return to later.

Next we consider the exact symmetry $K$. Since $K$ is a symmetry for each disorder realization $I$, $K$ is spontaneously broken if each realization breaks $K$, in the sense that $\langle O_e^{\dagger}(x)O_e(y)\rangle_I\to c_I\neq0$ as $|x-y|\to\infty$ for some operator $O_e$ charged under the exact symmetry $K$. The natural ensemble-averaged quantity that captures this sample-wise symmetry breaking order is the EA correlator 
\begin{equation}
\label{eq:EAcorrelator}
    C_e (|x-y|) = \overline{|\langle O_e(x)O^\dagger_e(y)\rangle|},
\end{equation}
which approaches a constant $c>0$ as $|x-y|\to\infty$ in an SSB phase. As for the first-moment correlator like Eq.~\eqref{eq:1stcorrelator} but with $O_e$, there are two possibilities: (1) $\overline{\langle O_e(x)O^{\dagger}_e(y)\rangle}\to c\neq0$, meaning that $K$ is also broken {as an average symmetry}, and (2) $\overline{\langle O_e(x)O^{\dagger}_e(y)\rangle}\to 0$, meaning that $K$ is unbroken {as an average symmetry}. The latter possibility is familiar in spin glass physics, where the spins order in each sample, but in random directions so that the symmetry still holds on average.

Assuming the absence of SSB order we can now motivate the other part of the power-law rule: the EA correlator, Eq.~\eqref{eq:EAcorrelator}, should decay as a power-law for some $O_e$ charged under the exact symmetry $K$. If the anomaly remains nontrivial even after reducing $K$ to an average symmetry, we would likewise require power-law decay for the first-moment correlator, Eq.~\eqref{eq:1stcorrelator}. However, in this work we focus on situations where $K$ must be an exact symmetry for the anomaly to remain nontrivial. In such cases, it is consistent for the first-moment correlator to decay exponentially, as we will see in later examples. Of course, there can be examples where the first-moment correlator decays critically even when this behavior is not required by an anomaly. In the language of quasi-long-range order, our result implies that the exact symmetry $K$ should at least be on the verge of spontaneously breaking down to an average symmetry.

We note that the relation between average anomaly and various correlation functions has been discussed from a different perspective (via mapping to mixed-state density matrices) in Refs.~\cite{XuJian2025, YouOshikawa}.\footnote{In that formulation, the disorder ensemble is represented by the mixed state $\rho_{\rm dis}=\overline{\rho_I}$, where $\rho_I$ is the ground-state density matrix of a disorder realization $I$, and the mixed anomaly between strong (exact) and weak (average) symmetries is captured by ordinary and R\'eny-$2$ correlation functions for operators charged under the weak and strong symmetries respectively. Here we address the complementary Hamiltonian problem of how the average anomaly constrains ground-state correlation functions in a gapless quenched ensemble. One important difference is that higher-moment correlation functions such as $\overline{|\langle O\rangle|}$ or $\overline{|\langle O\rangle|^2}$ are not encoded in the single density matrix $\overline{\rho_I}$; rather, they require higher-moment dencity matrices such as $\overline{\rho_I\otimes\rho_I}$. So our formulation in this work is not identical to earlier works from a mixed-state point of view.}

\section*{Examples}

\textbf{\boldmath Random Majorana lattice in $d=1,2$.} Our first example is the \textit{random Majorana hopping model} on bipartite lattices. Consider a chain in $1d$ or a square lattice in $2d$, with one Majorana fermion mode $\gamma_j$ on each lattice site $j$. The Hamiltonian is simply a nearest-neighbor hopping model with random hopping amplitude:
\begin{equation}
\label{eq:randomMajorana}
    H=-i\sum_{\langle jk\rangle}t_{jk}\gamma_j\gamma_k,
\end{equation}
with $t_{jk}=-t_{kj}$ an independent random real variable on each link $\langle jk\rangle$. This model preserves the average lattice translation symmetry as long as the probability distribution $P[t_{jk}]$ is translationally invariant. The microscopic LSM input is the unpaired Majorana mode in each unit cell; this leads to a nontrivial average anomaly involving exact fermion parity $Z^f_2$ and average translation $Z^d$~\cite{Hsieh_2016, Aksoy_2021}. Additionally, there is an exact time-reversal symmetry $\mathcal{T}: \gamma_j\to (-1)^{\sigma(j)}\gamma_j$, where $\sigma(j) = \pm 1$ depending the sublattice of $j$. This satisfies $\mathcal{T}^2 = +1$, placing the exact symmetry in class BDI~\cite{AltlandZirnbauer1997}. As a consequence of the anomaly, the ground state of the free-fermionic model is delocalized in any spatial dimension, even for $d=1,2$. Even in the presence of local interactions, the anomaly still requires the ground state to be either long-range entangled or to spontaneously break one of the symmetries involved in the anomaly. Since fermion parity cannot be spontaneously broken, the only remaining possibility is the spontaneous breaking of the average translation symmetry. However, spontaneous breaking of an average symmetry is forbidden in $d \le 2$ by the Imry-Ma constraint \cite{ImryMa}.

\begin{figure*}[t!]
     \centering
     \begin{subfigure}[t]{0.33\textwidth}
        \centering
        \caption{Random Majorana lattice, $d=1$}\includegraphics[width=0.95\linewidth]{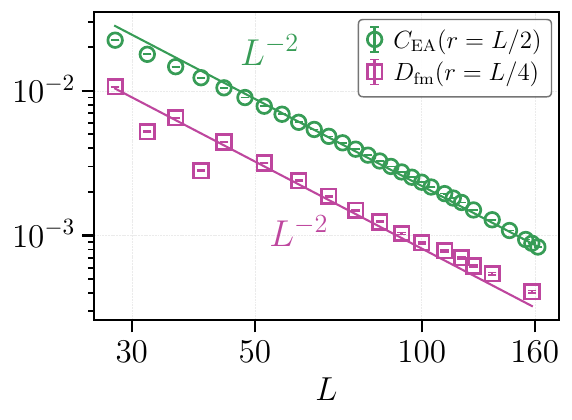}
        \label{Fig:1d-majorana-main}
    \end{subfigure}%
    \hfill
     \begin{subfigure}[t]{0.33\textwidth}
        \centering
        \caption{Random Majorana lattice, $d=2$}\includegraphics[width=0.95\linewidth]{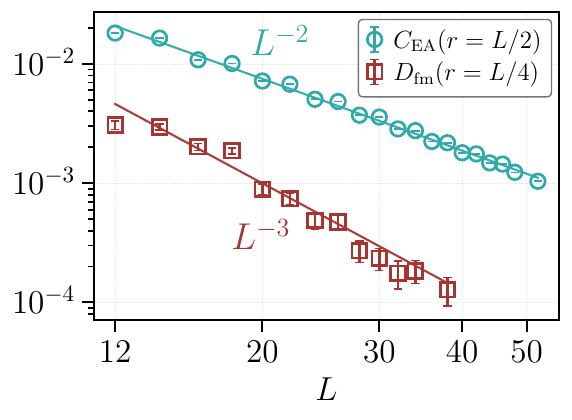}
        \label{Fig:2d-majorana-main}
    \end{subfigure}%
    \hfill
     \begin{subfigure}[t]{0.33\textwidth}
        \centering
        \caption{Random AFM Heisenberg chain}\includegraphics[width=0.95\linewidth]{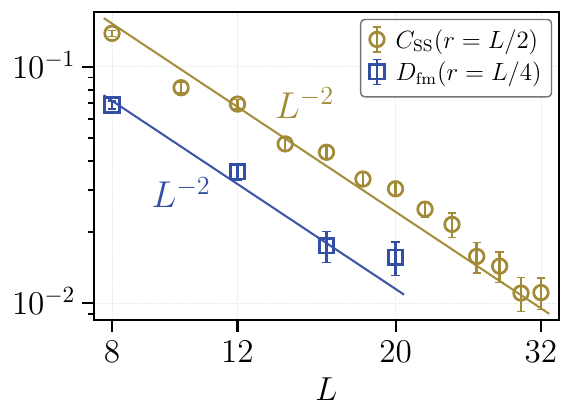}
        \label{Fig:AFM-heisenberg-main}
    \end{subfigure}%
    \caption{\justifying Power-law finite-size scaling of correlation functions in the random Majorana lattice model and the random antiferromagnetic Heisenberg chain, for operators charged under \textit{exact} symmetries ($C_\mathrm{EA}$ [Eq.~\ref{eq:EAcorrelator}] and $C_\mathrm{SS}$ [Eq.~\ref{eq:CSS}] respectively) and \textit{average} symmetries ($D_\mathrm{fm}$ [Eq.~\ref{eq:Dfm-def}] for both models). We only show data with signal-to-noise ratio (SNR) and lattice separation ($r$) above the thresholds stated below. (a) $1d$ random Majorana lattice: exact diagonalization with $10^{4}L$ disorder realizations; $r \ge 3$, $\mathrm{SNR} \ge 10$.(b) $2d$ random Majorana lattice: correlations evaluated at $r = r_x$, exact diagonalization with $200L$ disorder realizations; $r \ge 3$, $\mathrm{SNR} \ge 3$. (c) Random antiferromagnetic Heisenberg chain: DMRG, with sweeps performed until the energy change between successive sweeps was below $10^{-8}$, using $50$--$100$ disorder realizations; $r \ge 2$, $\mathrm{SNR} \ge 3$.}
    \label{Fig:main-powerlaw}
\end{figure*}

In $1d$, the model Eq.~\eqref{eq:randomMajorana} can be solved analytically using the strong-disorder renormalization group (SDRG) approach and is known to flow to an infinite-randomness fixed point akin to the random singlet state~\cite{Fisher1995Ising}. In $2d$, the model flows to a critical delocalized state known as the Gade phase~\cite{Gade1993,Gade1991}. It is also believed that the Gade phase will eventually flow to an infinite-randomness fixed point in the thermodynamic limit, although for numerically accessible system sizes its behavior is known to be more subtle~\cite{Motrunich2002}. In these strong-randomness settings, the power laws discussed below refer to disorder-averaged spatial correlators, and should not be confused with typical sample-wise correlators, which can exhibit activated scaling.

The $1d$ random Majorana model is closely related to several other strongly interacting models. Through a Jordan--Wigner transformation, the Majorana model can be mapped to a random transverse-field Ising model, which is solved using SDRG~\cite{Fisher1995Ising}. More explicitly, after pairing $\gamma_{2n-1}$ and $\gamma_{2n}$ into one Ising spin, the intra-pair hopping $i\gamma_{2n-1}\gamma_{2n}$ and the inter-pair hopping $i\gamma_{2n}\gamma_{2n+1}$ map, up to signs, to the random transverse-field and Ising-bond terms, respectively. One-site translation of the Majorana chain exchanges these two terms, and is therefore realized as Kramers--Wannier duality in the Ising chain. Under this Jordan--Wigner map, nearest-neighbor Majorana bilinears remain local, whereas single Majorana operators become nonlocal Ising order and disorder operators.
The random-bond antiferromagnetic Heisenberg chain can likewise be solved via SDRG~\cite{PhysRevB.50.3799}, exhibiting universal properties similar to those of the random Majorana model, which we comment on later. 

We now define the relevant correlation functions. The fermion operator $\gamma_j$ is charged under the exact fermion parity symmetry, so motivated by the power-law rule we study the EA correlator
\begin{equation}
    \label{eq:fermionEAcorrelator}
    C_{\mathrm{EA}}(r):=\frac{1}{L^d}\sum_j\overline{|\langle i\gamma_j\gamma_{j+r\hat{x}} \rangle|},
\end{equation}
where we average over positions $j$ ($L$ is the linear system size). We can also study the first-moment correlator $G(r)=\sum_j\overline{\langle i\gamma_j\gamma_{j+r}\rangle}/L^d$, which is not constrained by the anomaly. The numerical results are presented in Figs.~\ref{Fig:1d-majorana-main} and \ref{Fig:2d-majorana-main}.

We find that, in both $1d$ and $2d$, $C_{\mathrm{EA}}(r)$ indeed decays as a power-law, while $G(r)$ decays exponentially for generic disorder \cite{SM_RandomLesson}.
While the behaviors are fully consistent with the power-law rule from anomaly arguments, they are hardly surprising -- after all, even for an ordinary diffusive metal, the fermion Green's function $G(r)$ will decay exponentially beyond the mean free path, while the EA correlator Eq.~\eqref{eq:fermionEAcorrelator} has an algebraic diffusive tail, naturally estimated as $C^{diff}_{\mathrm{EA}} \sim 1/r^{(d+2)/2}$.\footnote{More precisely, in an ordinary diffusive metal, the corresponding second moment of equal-time fermion correlator satisfies
$\overline{|\langle c^{\dagger}_rc_0\rangle|^2}\sim 1/r^{d+2}$,  which follows from the standard density-density correlator $\overline{\langle \rho_r\rho_0\rangle}\sim Dq^2/(i\omega-Dq^2)$.~\cite{AndersonTransitionRMP}}

The real surprise comes when we consider first-moment correlation functions for operators charged under the \textit{average} translation symmetry. Let $d_j \equiv i\gamma_j \gamma_{j+\hat{x}}$ be a dimer operator. Its $k=\pi$ component,
$d_j^{\pi} = (-1)^{x_j} d_j$, is charged under the average translation symmetry, since
$\overline{\langle d_j^{\pi} \rangle}$ acquires a sign under translations along $\hat{x}$ for a
translation-invariant disorder ensemble. In the $1d$ Ising representation, $d_j^\pi$ is the local Kramers--Wannier-odd combination of the transverse-field and Ising-bond energy densities.
We compute the staggered first-moment correlator
\begin{equation}
D_{\mathrm{fm}}(r)
:=\frac{1}{L^d}\sum_j
\left(
\overline{\langle d_j d_{j+r} \rangle}
-
\overline{\langle d_j \rangle}
\,
\overline{\langle d_{j+r} \rangle}
\right)
(-1)^r,
\label{eq:Dfm-def}
\end{equation}
which is equivalently a first-moment correlator for $d^{\pi}$. Our numerical results are shown in Figs.~\ref{Fig:1d-majorana-main} and \ref{Fig:2d-majorana-main}. 
In $d=1$, we find a clear $1/r^{2}$ power-law decay, while in $d=2$ we observe a $1/r^{3}$ power-law decay. However, previous field-theoretic analysis suggests a continuous line of fixed points reminiscent of the Kosterlitz--Thouless transition~\cite{Konig2012}; thus we expect the power-law exponent to vary throughout the phase.
We note that Eq.~\eqref{eq:Dfm-def} differs from dimer correlators often studied in random-singlet settings~\cite{Shu_2016,PhysRevX.8.041040}, where the connected subtraction is performed before disorder averaging; those correlators are not constrained by the anomaly and exhibit different scaling, as discussed in the Supplemental Material~\cite{SM_RandomLesson}.

\vspace{5pt}
\textbf{\boldmath Random antiferromagnetic Heisenberg chain.} The power-law rule extends beyond free fermionic examples discussed so far, and holds for generic interacting systems. In the absence of analytically or numerically tractable interacting disordered systems in higher dimensions, we focus on the random antiferromagnetic (AFM) spin-$\frac{1}{2}$ Heisenberg chain~\cite{MaDasguptaHu1979, DasguptaMa1980}, 
\begin{equation}
    H = \sum_{i}J_{i,i+1}~S_i \cdot S_{i+1},
\end{equation}
which exhibits the well-known random singlet phase~\cite{PhysRevB.50.3799}.  This model has a spin-$1/2$ degree of freedom in each unit cell, with exact $SO(3)$ symmetry and an average translation symmetry, and thus realizes the average LSM anomaly~\cite{KimchiNahumSenthil, MaWang2023}. It has a power-law decaying spin-spin correlation~\cite{PhysRevB.50.3799}
\begin{equation}
C_{\mathrm{SS}}(r)
:= \frac{1}{L}\sum_{i}\overline{\big\langle \vec S_i \cdot \vec S_{i+r}\big\rangle}
\sim (-1)^{r}\, r^{-2},
\label{eq:CSS}
\end{equation}
consistent with the ``power-law rule" for the EA correlator\footnote{For random singlets, the standard spin-spin correlator is trivially also the EA correlator upto an overall sign, since $\langle S_i \cdot S_{j}\rangle = -3/4$, i.e. of uniform sign, if $i,j$ form a singlet, and zero otherwise. } which we also verify in Fig.~\ref{Fig:AFM-heisenberg-main}.
We further consider the dimer operator $d_i=\vec S_i\cdot \vec S_{i+1}$, whose staggered component is charged under the average translation symmetry, and find that the first-moment correlator defined in Eq.~\ref{eq:Dfm-def} decays as $1/r^{2}$ (Fig.~\ref{Fig:AFM-heisenberg-main}), exemplifying the power-law rule for average symmetries. This behavior can also be analytically explained using SDRG arguments~\cite{SM_RandomLesson}.

Finally, we expect this general phenomenon to also extend to higher dimensions and beyond-LSM average anomalies, which will be interesting to study in the future.
\section*{Experimental relevance}
A natural question regarding the power-law rule, which is especially relevant for its experimental significance, is whether these observables are self-averaging. To diagnose that, we studied the mean and standard error of each of these quantities at different $r, L$ (the separation and the system size, respectively), and study the corresponding signal-to-noise ratios (SNR). In the Supplementary
Material~\cite{SM_RandomLesson}, we provide evidence that the SNRs for these quantities satisfy an \mbox{asymptotic power-law behavior,}
\begin{align}
    \mathrm{SNR} \sim L^ a/ r^b, ~~\text{with}~~ b > a > 0,
\end{align}
which suggests that these correlators are self-averaging in the thermodynamic setting where $r$ is fixed and $L\to \infty$.

We highlight that these correlators are experimentally accessible in both synthetic quantum matter (such as cold atomic gases in optical lattices), and in solid-state quantum materials. In cold atomic platforms, where snapshots of disordered ground states can be collected, higher moment correlators of Edwards--Anderson type as well as regular correlation functions can be measured to diagnose the anomaly.

However, in quantum materials, EA correlators are generally inaccessible.  Experimental scattering probes such as inelastic neutron scattering {and X-ray scattering} typically measure dynamical structure factors
$S_{O}(\mathbf{q},\omega)$, which are space-time Fourier transforms of two-point correlation functions of suitable local operators
$O_{\bf{r}}$. In a given sample and for an operator $O_{\bf{r}}$ the structure factor may be written as
\begin{equation}
S_{O}(\mathbf{q},\omega)
= \sum_{\mathbf{r}} \int_{-\infty}^{\infty} dt\;
e^{i(\omega t - \mathbf{q}\cdot\mathbf{r})}\,
\langle O_{\mathbf{r}}(t)\,O_{\mathbf{0}}(0)\rangle.
\label{eq:structurefactor}
\end{equation}
Assuming the correlation function self averages (which we checked for all the examples we studied), the expectation value can be replaced by the sample-averaged one $\langle\cdots \rangle \to \overline{\langle\cdots \rangle}$. Now, the equal-time structure factor $S_{O}(\mathbf{q})=\int \frac{d\omega}{2\pi}\,S_{O}(\mathbf{q},\omega)$ is directly related to the disorder-averaged \textit{first-moment} spatial correlator, since $\overline{\langle O_{\mathbf{r}}(t)\rangle} = 0$ for charged operators as an ensemble consequence of the average symmetry, although $\langle O_{\mathbf{r}}(t)\rangle_I$ need not vanish in a fixed disorder realization $I$. Hence, such disorder-averaged first-moment two-point functions are experimentally accessible. 
For random-singlet and related critical chains at strong randomness, the spin-spin structure factor has been studied using strong-disorder RG ~\cite{Damle2000,Motrunich2001}, numerically~\cite{Shu2018}, and experimentally through indirect probes in~\cite{Shiroka2011}. In two dimensions, scattering studies of disordered quantum spin-liquid (QSL) candidates such as YbMgGaO$_4$~\cite{Li_2016_PRL_muSR_YbMgGaO4, Paddison_2017_NatPhys, Shen_2016_Nature, Li_2017_NatCommun} reveal QSL phenomenology consistent with the possibility of higher dimensional analogues of long-range valence bond correlations~\cite{KimchiNahumSenthil}. Model Hamiltonians realizing random-singlet phases in $2d$ have also been identified numerically~\cite{PhysRevX.8.041040}. It will be interesting to design experiments in such materials to probe the anomaly-enforced critical correlations uncovered in our work. In a strongly disordered quantum-critical system, suitable choices of operators $O_i$ that transform nontrivially under the anomalous average global symmetry (such as dimer operators in the examples studied here) may allow direct observation of the predicted nontrivial power-law decay of disorder-averaged spatial correlations via the corresponding $S_{O}(\mathbf{q},\omega)$ or $S_{O}(\mathbf{q})$.  Notably, the dimer operators being spinless, should couple to charge density operators that are probed by X-ray scattering.

\textbf{Acknowledgements}: We thank Francisco Divi and Jinmin Yi for illuminating discussions. Y.P. acknowledges support from the Natural Sciences and Engineering Research Council of Canada (NSERC) through Discovery Grants. Research at Perimeter Institute is supported in part by the Government of Canada through the Department of Innovation, Science and Economic Development; and by the Province of Ontario through the Ministry of Colleges, Universities, Research Excellence and Security. Part of this work was done at the Kavli Institute for Theoretical Physics during the ``Noise-robust Phases of Quantum Matter'' program, which was supported in part by grant NSF PHY-2309135 to the Kavli Institute for Theoretical Physics (KITP).

\bibliography{references}

@ARTICLE{YouOshikawa,
       author = {{You}, Yizhi and {Oshikawa}, Masaki},
        title = "{Intrinsic symmetry-protected topological mixed state from modulated symmetries and hierarchical structure of boundary anomaly}",
      journal = {\prb},
         year = 2024,
        month = oct,
       volume = {110},
       number = {16},
          eid = {165160},
        pages = {165160},
          doi = {10.1103/PhysRevB.110.165160},
archivePrefix = {arXiv},
       eprint = {2407.08786},
 primaryClass = {quant-ph},
       adsurl = {https://ui.adsabs.harvard.edu/abs/2024PhRvB.110p5160Y}
}

@ARTICLE{CalveraWang2021,
       author = {{Calvera}, Vladimir and {Wang}, Chong},
        title = "{Theory of Dirac Spin-Orbital Liquids: monopoles, anomalies, and applications to $SU(4)$ honeycomb models}",
      journal = {arXiv e-prints},
         year = 2021,
        month = mar,
          eid = {arXiv:2103.13405},
        pages = {arXiv:2103.13405},
          doi = {10.48550/arXiv.2103.13405},
archivePrefix = {arXiv},
       eprint = {2103.13405},
 primaryClass = {cond-mat.str-el},
       adsurl = {https://ui.adsabs.harvard.edu/abs/2021arXiv210313405C}
}

@ARTICLE{ZhangSongSenthil2025,
       author = {{Zhang}, Yunchao and {Song}, Xue-Yang and {Senthil}, T.},
        title = "{Unnecessary quantum criticality in $SU(3)$ kagome magnets}",
      journal = {arXiv e-prints},
         year = 2025,
        month = aug,
          eid = {arXiv:2508.16725},
        pages = {arXiv:2508.16725},
          doi = {10.48550/arXiv.2508.16725},
archivePrefix = {arXiv},
       eprint = {2508.16725},
 primaryClass = {cond-mat.str-el},
       adsurl = {https://ui.adsabs.harvard.edu/abs/2025arXiv250816725Z}
}

@ARTICLE{Metlitski2018,
       author = {{Metlitski}, Max A. and {Thorngren}, Ryan},
        title = "{Intrinsic and emergent anomalies at deconfined critical points}",
      journal = {\prb},
         year = 2018,
        month = aug,
       volume = {98},
       number = {8},
          eid = {085140},
        pages = {085140},
          doi = {10.1103/PhysRevB.98.085140},
archivePrefix = {arXiv},
       eprint = {1707.07686},
 primaryClass = {cond-mat.str-el},
       adsurl = {https://ui.adsabs.harvard.edu/abs/2018PhRvB..98h5140M}
}

@ARTICLE{Else2020,
       author = {{Else}, Dominic V. and {Thorngren}, Ryan},
        title = "{Topological theory of Lieb-Schultz-Mattis theorems in quantum spin systems}",
      journal = {\prb},
         year = 2020,
        month = jun,
       volume = {101},
       number = {22},
          eid = {224437},
        pages = {224437},
          doi = {10.1103/PhysRevB.101.224437},
archivePrefix = {arXiv},
       eprint = {1907.08204},
 primaryClass = {cond-mat.str-el},
       adsurl = {https://ui.adsabs.harvard.edu/abs/2020PhRvB.101v4437E}
}

@ARTICLE{Po2017,
       author = {{Po}, Hoi Chun and {Watanabe}, Haruki and {Jian}, Chao-Ming and {Zaletel}, Michael P.},
        title = "{Lattice Homotopy Constraints on Phases of Quantum Magnets}",
      journal = {\prl},
         year = 2017,
        month = sep,
       volume = {119},
       number = {12},
          eid = {127202},
        pages = {127202},
          doi = {10.1103/PhysRevLett.119.127202},
archivePrefix = {arXiv},
       eprint = {1703.06882},
 primaryClass = {cond-mat.str-el},
       adsurl = {https://ui.adsabs.harvard.edu/abs/2017PhRvL.119l7202P}
}

@ARTICLE{ElseThorngrenSenthil2020,
       author = {{Else}, Dominic V. and {Thorngren}, Ryan and {Senthil}, T.},
        title = "{Non-Fermi Liquids as Ersatz Fermi Liquids: General Constraints on Compressible Metals}",
      journal = {Physical Review X},
         year = 2021,
        month = apr,
       volume = {11},
       number = {2},
          eid = {021005},
        pages = {021005},
          doi = {10.1103/PhysRevX.11.021005},
archivePrefix = {arXiv},
       eprint = {2007.07896},
 primaryClass = {cond-mat.str-el},
       adsurl = {https://ui.adsabs.harvard.edu/abs/2021PhRvX..11b1005E}
}

@article{else_classifying_2014,
  title = {Classifying Symmetry-Protected Topological Phases through the Anomalous Action of the Symmetry on the Edge},
  author = {Else, Dominic V. and Nayak, Chetan},
  year = {2014},
  month = dec,
  journal = {Physical Review B},
  volume = {90},
  number = {23},
  pages = {235137},
  publisher = {{American Physical Society}},
  doi = {10.1103/PhysRevB.90.235137}
}

@ARTICLE{Zou2021,
       author = {{Zou}, Liujun and {He}, Yin-Chen and {Wang}, Chong},
        title = "{Stiefel Liquids: Possible Non-Lagrangian Quantum Criticality from Intertwined Orders}",
      journal = {Physical Review X},
         year = 2021,
        month = jul,
       volume = {11},
       number = {3},
          eid = {031043},
        pages = {031043},
          doi = {10.1103/PhysRevX.11.031043},
archivePrefix = {arXiv},
       eprint = {2101.07805},
 primaryClass = {cond-mat.str-el},
       adsurl = {https://ui.adsabs.harvard.edu/abs/2021PhRvX..11c1043Z}
}

@ARTICLE{LYZZ2025,
       author = {{Liu}, Ruizhi and {Yi}, Jinmin and {Zhou}, Shiyu and {Zou}, Liujun},
        title = "{Entanglement area law and Lieb-Schultz-Mattis theorem in long-range interacting systems, and symmetry-enforced long-range entanglement}",
      journal = {\prb},
         year = 2025,
        month = dec,
       volume = {112},
       number = {21},
          eid = {214408},
        pages = {214408},
          doi = {10.1103/2jgh-nrj1},
archivePrefix = {arXiv},
       eprint = {2405.14929},
 primaryClass = {cond-mat.str-el},
       adsurl = {https://ui.adsabs.harvard.edu/abs/2025PhRvB.112u4408L}
}

@ARTICLE{GioiaWangMomentum,
       author = {{Gioia}, Lei and {Wang}, Chong},
        title = "{Nonzero Momentum Requires Long-Range Entanglement}",
      journal = {Physical Review X},
         year = 2022,
        month = jul,
       volume = {12},
       number = {3},
          eid = {031007},
        pages = {031007},
          doi = {10.1103/PhysRevX.12.031007},
archivePrefix = {arXiv},
       eprint = {2112.06946},
 primaryClass = {cond-mat.str-el},
       adsurl = {https://ui.adsabs.harvard.edu/abs/2022PhRvX..12c1007G}
}

@ARTICLE{ChenGuWenLRE,
       author = {{Chen}, Xie and {Gu}, Zheng-Cheng and {Wen}, Xiao-Gang},
        title = "{Local unitary transformation, long-range quantum entanglement, wave function renormalization, and topological order}",
      journal = {\prb},
         year = 2010,
        month = oct,
       volume = {82},
       number = {15},
          eid = {155138},
        pages = {155138},
          doi = {10.1103/PhysRevB.82.155138},
archivePrefix = {arXiv},
       eprint = {1004.3835},
 primaryClass = {cond-mat.str-el},
       adsurl = {https://ui.adsabs.harvard.edu/abs/2010PhRvB..82o5138C}
}

@ARTICLE{AndersonTransitionRMP,
       author = {{Evers}, Ferdinand and {Mirlin}, Alexander D.},
        title = "{Anderson transitions}",
      journal = {Reviews of Modern Physics},
         year = 2008,
        month = oct,
       volume = {80},
       number = {4},
        pages = {1355-1417},
          doi = {10.1103/RevModPhys.80.1355},
archivePrefix = {arXiv},
       eprint = {0707.4378},
 primaryClass = {cond-mat.mes-hall},
       adsurl = {https://ui.adsabs.harvard.edu/abs/2008RvMP...80.1355E}
}

@ARTICLE{XuJian2025,
       author = {{Xu}, Yichen and {Jian}, Chao-Ming},
        title = "{Average-exact mixed anomalies and compatible phases}",
      journal = {\prb},
         year = 2025,
        month = mar,
       volume = {111},
       number = {12},
          eid = {125128},
        pages = {125128},
          doi = {10.1103/PhysRevB.111.125128},
archivePrefix = {arXiv},
       eprint = {2406.07417},
 primaryClass = {cond-mat.str-el},
       adsurl = {https://ui.adsabs.harvard.edu/abs/2025PhRvB.111l5128X}
}

@ARTICLE{Isingcorrelation,
       author = {{Igl{\'o}i}, Ferenc and {Kov{\'a}cs}, Istv{\'a}n A.},
        title = "{Transverse spin correlations of the random transverse-field Ising model}",
      journal = {\prb},
         year = 2018,
        month = mar,
       volume = {97},
       number = {9},
          eid = {094205},
        pages = {094205},
          doi = {10.1103/PhysRevB.97.094205},
archivePrefix = {arXiv},
       eprint = {1712.07467},
 primaryClass = {cond-mat.dis-nn},
       adsurl = {https://ui.adsabs.harvard.edu/abs/2018PhRvB..97i4205I}
}

@book{AltlandBook,
  author = {Altland, Alexander and Simons, Ben D},
  publisher = {Cambridge University Press},
  title = {Condensed matter field theory},
  year = 2010
}

@article{ImryMa,
  title = {Random-Field Instability of the Ordered State of Continuous Symmetry},
  author = {Imry, Yoseph and Ma, Shang-keng},
  journal = {Phys. Rev. Lett.},
  volume = {35},
  issue = {21},
  pages = {1399--1401},
  numpages = {0},
  year = {1975},
  month = {Nov},
  publisher = {American Physical Society},
  doi = {10.1103/PhysRevLett.35.1399},
  url = {https://link.aps.org/doi/10.1103/PhysRevLett.35.1399}
}

@ARTICLE{KimchiNahumSenthil,
       author = {{Kimchi}, Itamar and {Nahum}, Adam and {Senthil}, T.},
        title = "{Valence Bonds in Random Quantum Magnets: Theory and Application to YbMgGaO$_{4}$}",
      journal = {Physical Review X},
         year = 2018,
        month = jul,
       volume = {8},
       number = {3},
          eid = {031028},
        pages = {031028},
          doi = {10.1103/PhysRevX.8.031028},
archivePrefix = {arXiv},
       eprint = {1710.06860},
 primaryClass = {cond-mat.str-el},
       adsurl = {https://ui.adsabs.harvard.edu/abs/2018PhRvX...8c1028K}
}

@ARTICLE{MZBCW2025,
       author = {{Ma}, Ruochen and {Zhang}, Jian-Hao and {Bi}, Zhen and {Cheng}, Meng and {Wang}, Chong},
        title = "{Topological Phases with Average Symmetries: The Decohered, the Disordered, and the Intrinsic}",
      journal = {Physical Review X},
         year = 2025,
        month = apr,
       volume = {15},
       number = {2},
          eid = {021062},
        pages = {021062},
          doi = {10.1103/PhysRevX.15.021062},
archivePrefix = {arXiv},
       eprint = {2305.16399},
 primaryClass = {cond-mat.str-el},
       adsurl = {https://ui.adsabs.harvard.edu/abs/2025PhRvX..15b1062M}
}

@ARTICLE{MaWang2023,
       author = {{Ma}, Ruochen and {Wang}, Chong},
        title = "{Average Symmetry-Protected Topological Phases}",
      journal = {Physical Review X},
         year = 2023,
        month = jul,
       volume = {13},
       number = {3},
          eid = {031016},
        pages = {031016},
          doi = {10.1103/PhysRevX.13.031016},
archivePrefix = {arXiv},
       eprint = {2209.02723},
 primaryClass = {cond-mat.str-el},
       adsurl = {https://ui.adsabs.harvard.edu/abs/2023PhRvX..13c1016M}
}

@article{PhysRevB.50.3799,
  title = {Random antiferromagnetic quantum spin chains},
  author = {Fisher, Daniel S.},
  journal = {Phys. Rev. B},
  volume = {50},
  issue = {6},
  pages = {3799--3821},
  numpages = {0},
  year = {1994},
  month = {Aug},
  publisher = {American Physical Society},
  doi = {10.1103/PhysRevB.50.3799},
  url = {https://link.aps.org/doi/10.1103/PhysRevB.50.3799}
}

@article{Shu_2016,
   title={Properties of the random-singlet phase: From the disordered Heisenberg chain to an amorphous valence-bond solid},
   volume={94},
   ISSN={2469-9969},
   url={http://dx.doi.org/10.1103/PhysRevB.94.174442},
   DOI={10.1103/physrevb.94.174442},
   number={17},
   journal={Physical Review B},
   publisher={American Physical Society (APS)},
   author={Shu, Yu-Rong and Yao, Dao-Xin and Ke, Chih-Wei and Lin, Yu-Cheng and Sandvik, Anders W.},
   year={2016},
   month=nov }

@article{LiebSchultzMattis1961,
  title        = {Two soluble models of an antiferromagnetic chain},
  author       = {Lieb, Elliott and Schultz, Theodore and Mattis, Daniel},
  journal      = {Annals of Physics},
  volume       = {16},
  pages        = {407--466},
  year         = {1961},
  doi          = {10.1016/0003-4916(61)90115-4}
}

@ARTICLE{Oshikawa1999LSM,
       author = {{Oshikawa}, Masaki},
        title = "{Commensurability, Excitation Gap, and Topology in Quantum Many-Particle Systems on a Periodic Lattice}",
      journal = {\prl},
         year = "2000",
        month = "Feb",
       volume = {84},
       number = {7},
        pages = {1535-1538},
          doi = {10.1103/PhysRevLett.84.1535},
archivePrefix = {arXiv},
       eprint = {cond-mat/9911137},
 primaryClass = {cond-mat.str-el},
       adsurl = {https://ui.adsabs.harvard.edu/abs/2000PhRvL..84.1535O}
}

@article{Oshikawa2000Luttinger,
  title        = {Topological approach to Luttinger's theorem and the Fermi surface of a Kondo lattice},
  author       = {Oshikawa, Masaki},
  journal      = {Physical Review Letters},
  volume       = {84},
  pages        = {3370--3373},
  year         = {2000},
  doi          = {10.1103/PhysRevLett.84.3370},
  eprint       = {cond-mat/0002392},
  archivePrefix= {arXiv}
}

@article{Hastings2004LSM,
  title        = {Lieb-Schultz-Mattis in higher dimensions},
  author       = {Hastings, M. B.},
  journal      = {Physical Review B},
  volume       = {69},
  pages        = {104431},
  year         = {2004},
  doi          = {10.1103/PhysRevB.69.104431},
  eprint       = {cond-mat/0305505},
  archivePrefix= {arXiv}
}

@article{Gade1993,
title = {Anderson localization for sublattice models},
journal = {Nuclear Physics B},
volume = {398},
number = {3},
pages = {499-515},
year = {1993},
issn = {0550-3213},
doi = {https://doi.org/10.1016/0550-3213(93)90601-K},
url = {https://www.sciencedirect.com/science/article/pii/055032139390601K},
author = {Renate Gade},
}

@article{Gade1991,
title = {The n = 0 replica limit of U(n) and U(n)SO(n) models},
journal = {Nuclear Physics B},
volume = {360},
number = {2},
pages = {213-218},
year = {1991},
issn = {0550-3213},
doi = {https://doi.org/10.1016/0550-3213(91)90401-I},
url = {https://www.sciencedirect.com/science/article/pii/055032139190401I},
author = {Renate Gade and Franz Wegner},
}

@misc{SM_RandomLesson,
  author = {},
  title = {Please see the {Supplemental Material} for further discussion of replica-theory derivations of anomaly-constrained correlations, strong-disorder RG analyses of first-moment correlators in the $1d$ random-singlet-type phases, additional numerical results for representative disordered models, and the noise and self-averaging behavior of the correlators.} 
}

@article{Lanzetta_2023,
   title={Bootstrapping Lieb-Schultz-Mattis anomalies},
   volume={107},
   ISSN={2469-9969},
   url={http://dx.doi.org/10.1103/PhysRevB.107.205137},
   DOI={10.1103/physrevb.107.205137},
   number={20},
   journal={Physical Review B},
   publisher={American Physical Society (APS)},
   author={Lanzetta, Ryan A. and Fidkowski, Lukasz},
   year={2023},
   month=may }

@article{Cheng2016LSM,
   title={Translational Symmetry and Microscopic Constraints on Symmetry-Enriched Topological Phases: A View from the Surface},
   volume={6},
   ISSN={2160-3308},
   url={http://dx.doi.org/10.1103/PhysRevX.6.041068},
   DOI={10.1103/physrevx.6.041068},
   number={4},
   journal={Physical Review X},
   publisher={American Physical Society (APS)},
   author={Cheng, Meng and Zaletel, Michael and Barkeshli, Maissam and Vishwanath, Ashvin and Bonderson, Parsa},
   year={2016},
   month=dec }

@article{Cho2017LSM,
   title={Anomaly manifestation of Lieb-Schultz-Mattis theorem and topological phases},
   volume={96},
   ISSN={2469-9969},
   url={http://dx.doi.org/10.1103/PhysRevB.96.195105},
   DOI={10.1103/physrevb.96.195105},
   number={19},
   journal={Physical Review B},
   publisher={American Physical Society (APS)},
   author={Cho, Gil Young and Hsieh, Chang-Tse and Ryu, Shinsei},
   year={2017},
   month=nov }

@article{Cheng2023LSM,
   title={Lieb-Schultz-Mattis, Luttinger, and 't Hooft - anomaly matching in lattice systems},
   volume={15},
   ISSN={2542-4653},
   url={http://dx.doi.org/10.21468/SciPostPhys.15.2.051},
   DOI={10.21468/scipostphys.15.2.051},
   number={2},
   journal={SciPost Physics},
   publisher={Stichting SciPost},
   author={Cheng, Meng and Seiberg, Nathan},
   year={2023},
   month=aug }

@article{Seiberg2025LSM,
   title={LSM and CPT},
   volume={2025},
   ISSN={1029-8479},
   url={http://dx.doi.org/10.1007/JHEP11(2025)116},
   DOI={10.1007/jhep11(2025)116},
   number={11},
   journal={Journal of High Energy Physics},
   publisher={Springer Science and Business Media LLC},
   author={Seiberg, Nathan and Shao, Shu-Heng and Zhang, Wucheng},
   year={2025},
   month=nov }

@misc{tasaki2022LSMreview,
      title={The Lieb-Schultz-Mattis Theorem: A Topological Point of View}, 
      author={Hal Tasaki},
      year={2022},
      eprint={2202.06243},
      archivePrefix={arXiv},
      primaryClass={cond-mat.stat-mech},
      url={https://arxiv.org/abs/2202.06243}, 
}

@article{Shu2018,
  author  = {Shu, Yu-Rong and Dupont, Maxime and Yao, Dao-Xin and Capponi, Sylvain and Sandvik, Anders W.},
  title   = {Dynamical properties of the $S=\frac{1}{2}$ random Heisenberg chain},
  journal = {Physical Review B},
  year    = {2018},
  volume  = {97},
  pages   = {104424},
  doi     = {10.1103/PhysRevB.97.104424},
}

@article{Damle2000,
  author  = {Damle, Kedar and Motrunich, Olexei and Huse, David A.},
  title   = {Dynamics and transport in random antiferromagnetic spin chains},
  journal = {Physical Review Letters},
  year    = {2000},
  volume  = {84},
  number  = {15},
  pages   = {3434--3437},
  doi     = {10.1103/PhysRevLett.84.3434},
}

@article{Motrunich2001,
  author  = {Motrunich, Olexei and Damle, Kedar and Huse, David A.},
  title   = {Dynamics and transport in random quantum systems governed by strong-randomness fixed points},
  journal = {Physical Review B},
  year    = {2001},
  volume  = {63},
  pages   = {134424},
  doi     = {10.1103/PhysRevB.63.134424},
}

@article{Jian2018lsm,
  title = {Lieb-Schultz-Mattis theorem and its generalizations from the perspective of the symmetry-protected topological phase},
  author = {Jian, Chao-Ming and Bi, Zhen and Xu, Cenke},
  journal = {Phys. Rev. B},
  volume = {97},
  issue = {5},
  pages = {054412},
  numpages = {12},
  year = {2018},
  month = {Feb},
  publisher = {American Physical Society},
  doi = {10.1103/PhysRevB.97.054412},
  url = {https://link.aps.org/doi/10.1103/PhysRevB.97.054412}
}

@article{Shiroka2011,
  author  = {Shiroka, T. and Casola, F. and Glazkov, V. and Zheludev, A. and Prsa, K. and Ott, H.-R. and Mesot, J.},
  title   = {Distribution of NMR relaxations in a random Heisenberg chain},
  journal = {Physical Review Letters},
  year    = {2011},
  volume  = {106},
  pages   = {137202},
  doi     = {10.1103/PhysRevLett.106.137202},
}

@article{tHooft1979,
    author = "'t Hooft, Gerard",
    editor = "'t Hooft, Gerard and Itzykson, C. and Jaffe, A. and Lehmann, H. and Mitter, P. K. and Singer, I. M. and Stora, R.",
    title = "{Naturalness, chiral symmetry, and spontaneous chiral symmetry breaking}",
    reportNumber = "PRINT-80-0083 (UTRECHT)",
    doi = "10.1007/978-1-4684-7571-5_9",
    journal = "NATO Sci. Ser. B",
    volume = "59",
    pages = "135--157",
    year = "1980"
}

@article{MaDasguptaHu1979,
  title = {Random Antiferromagnetic Chain},
  author = {Ma, Shang-keng and Dasgupta, Chandan and Hu, Chin-kun},
  journal = {Phys. Rev. Lett.},
  volume = {43},
  issue = {19},
  pages = {1434--1437},
  numpages = {0},
  year = {1979},
  month = {Nov},
  publisher = {American Physical Society},
  doi = {10.1103/PhysRevLett.43.1434},
  url = {https://link.aps.org/doi/10.1103/PhysRevLett.43.1434}
}

@article{DasguptaMa1980,
  title = {Low-temperature properties of the random Heisenberg antiferromagnetic chain},
  author = {Dasgupta, Chandan and Ma, Shang-keng},
  journal = {Phys. Rev. B},
  volume = {22},
  issue = {3},
  pages = {1305--1319},
  numpages = {0},
  year = {1980},
  month = {Aug},
  publisher = {American Physical Society},
  doi = {10.1103/PhysRevB.22.1305},
  url = {https://link.aps.org/doi/10.1103/PhysRevB.22.1305}
}

@article{Igloi2018,
  author       = {Igl{\'o}i, Ferenc and Monthus, C{\'e}cile},
  title        = {Strong disorder {RG} approach -- a short review of recent developments},
  journal      = {The European Physical Journal B},
  year         = {2018},
  volume       = {91},
  number       = {11},
  pages        = {290},
  doi          = {10.1140/epjb/e2018-90434-8},
  url          = {https://doi.org/10.1140/epjb/e2018-90434-8},
  date         = {2018-11-21}
}

@article{PhysRevB.82.054437,
  title = {Renormalization group study of the two-dimensional random transverse-field Ising model},
  author = {Kov\'acs, Istv\'an A. and Igl\'oi, Ferenc},
  journal = {Phys. Rev. B},
  volume = {82},
  issue = {5},
  pages = {054437},
  numpages = {13},
  year = {2010},
  month = {Aug},
  publisher = {American Physical Society},
  doi = {10.1103/PhysRevB.82.054437},
  url = {https://link.aps.org/doi/10.1103/PhysRevB.82.054437}
}

@article{PhysRevX.8.041040,
  title = {Random-Singlet Phase in Disordered Two-Dimensional Quantum Magnets},
  author = {Liu, Lu and Shao, Hui and Lin, Yu-Cheng and Guo, Wenan and Sandvik, Anders W.},
  journal = {Phys. Rev. X},
  volume = {8},
  issue = {4},
  pages = {041040},
  numpages = {33},
  year = {2018},
  month = {Dec},
  publisher = {American Physical Society},
  doi = {10.1103/PhysRevX.8.041040},
  url = {https://link.aps.org/doi/10.1103/PhysRevX.8.041040}
}

@article{PhysRevB.108.064201,
  title = {Random geometry at an infinite-randomness fixed point},
  author = {Pandey, Akshat and Mahadevan, Aditya and Cowsik, Aditya},
  journal = {Phys. Rev. B},
  volume = {108},
  issue = {6},
  pages = {064201},
  numpages = {17},
  year = {2023},
  month = {Aug},
  publisher = {American Physical Society},
  doi = {10.1103/PhysRevB.108.064201},
  url = {https://link.aps.org/doi/10.1103/PhysRevB.108.064201}
}

@article{Hsieh_2016,
   title={All Majorana Models with Translation Symmetry are Supersymmetric},
   volume={117},
   ISSN={1079-7114},
   url={http://dx.doi.org/10.1103/PhysRevLett.117.166802},
   DOI={10.1103/physrevlett.117.166802},
   number={16},
   journal={Physical Review Letters},
   publisher={American Physical Society (APS)},
   author={Hsieh, Timothy H. and Hal\'asz, Gábor B. and Grover, Tarun},
   year={2016},
   month=oct }

@article{Aksoy_2021,
   title={Lieb-Schultz-Mattis type theorems for Majorana models with discrete symmetries},
   volume={104},
   ISSN={2469-9969},
   url={http://dx.doi.org/10.1103/PhysRevB.104.075146},
   DOI={10.1103/physrevb.104.075146},
   number={7},
   journal={Physical Review B},
   publisher={American Physical Society (APS)},
   author={Aksoy, O\:mer M. and Tiwari, Apoorv and Mudry, Christopher},
   year={2021},
   month=aug }

@article{Ludwig1994,
  author  = {Ludwig, A. W. W. and Fisher, M. P. A. and Shankar, R. and Grinstein, G.},
  title   = {{Integer} quantum {Hall} transition: An alternative approach and exact results},
  journal = {Physical Review B},
  volume  = {50},
  number  = {11},
  pages   = {7526--7552},
  year    = {1994},
  doi     = {10.1103/PhysRevB.50.7526}
}

@article{Motrunich2002,
  author  = {Motrunich, O. and Damle, K. and Huse, D. A.},
  title   = {Particle-hole symmetric localization in two dimensions},
  journal = {Physical Review B},
  volume  = {65},
  pages   = {064206},
  year    = {2002},
  doi     = {10.1103/PhysRevB.65.064206}
}

@article{Mudry2003,
  author  = {Mudry, C. and Ryu, S. and Furusaki, A.},
  title   = {Density of states for the {$\pi$}-flux state with bipartite real random hopping only: A weak disorder approach},
  journal = {Physical Review B},
  volume  = {67},
  pages   = {064202},
  year    = {2003},
  doi     = {10.1103/PhysRevB.67.064202}
}

@article{Hafner2014,
  author  = {H{\"a}fner, V. and Schindler, J. and Weik, N. and Mayer, T. and Balakrishnan, S. and Narayanan, R. and Bera, S. and Evers, F.},
  title   = {Density of States in Graphene with Vacancies: Midgap Power Law and Frozen Multifractality},
  journal = {Physical Review Letters},
  volume  = {113},
  pages   = {186802},
  year    = {2014},
  doi     = {10.1103/PhysRevLett.113.186802}
}

@article{Sanyal2016,
  author  = {Sanyal, S. and Damle, K. and Motrunich, O. I.},
  title   = {Vacancy-Induced Low-Energy States in Undoped Graphene},
  journal = {Physical Review Letters},
  volume  = {117},
  pages   = {116806},
  year    = {2016},
  doi     = {10.1103/PhysRevLett.117.116806}
}

@article{Konig2012,
  author  = {K{\"o}nig, E. J. and Ostrovsky, P. M. and Protopopov, I. V. and Mirlin, A. D.},
  title   = {Metal-insulator transition in two-dimensional random fermion systems of chiral symmetry classes},
  journal = {Physical Review B},
  volume  = {85},
  pages   = {195130},
  year    = {2012},
  doi     = {10.1103/PhysRevB.85.195130}
}

@article{Karcher2023,
  author  = {Karcher, J. F. and Gruzberg, I. A. and Mirlin, A. D.},
  title   = {Metal-insulator transition in a two-dimensional system of chiral unitary class},
  journal = {Physical Review B},
  volume  = {107},
  pages   = {L020201},
  year    = {2023},
  doi     = {10.1103/PhysRevB.107.L020201}
}

@article{Nayak2024,
  author  = {Nayak, N. P. and Sarkar, S. and Damle, K. and Bera, S.},
  title   = {Band-center metal-insulator transition in bond-disordered graphene},
  journal = {Physical Review B},
  volume  = {109},
  pages   = {035109},
  year    = {2024},
  doi     = {10.1103/PhysRevB.109.035109}
}

@article{Das2025Z2FluxDisorder,
  author  = {Das, Hiranmay and Nayak, Naba P. and Bera, Soumya and Shenoy, Vijay B.},
  title   = {Critical States of Fermions with $\mathbb{Z}_2$ Flux Disorder},
  journal = {Physical Review Letters},
  volume  = {135},
  number  = {25},
  pages   = {256305},
  year    = {2025},
  month   = dec,
  doi     = {10.1103/wsqw-2xh4},
  publisher = {American Physical Society}
}

@article{AltlandZirnbauer1997,
  title = {Nonstandard symmetry classes in mesoscopic normal-superconducting hybrid structures},
  author = {Altland, Alexander and Zirnbauer, Martin R.},
  journal = {Phys. Rev. B},
  volume = {55},
  issue = {2},
  pages = {1142--1161},
  numpages = {0},
  year = {1997},
  month = {Jan},
  publisher = {American Physical Society},
  doi = {10.1103/PhysRevB.55.1142},
  url = {https://link.aps.org/doi/10.1103/PhysRevB.55.1142}
}

@article{Fisher1995Ising,
  title = {Critical behavior of random transverse-field Ising spin chains},
  author = {Fisher, Daniel S.},
  journal = {Phys. Rev. B},
  volume = {51},
  issue = {10},
  pages = {6411--6461},
  numpages = {0},
  year = {1995},
  month = {Mar},
  publisher = {American Physical Society},
  doi = {10.1103/PhysRevB.51.6411},
  url = {https://link.aps.org/doi/10.1103/PhysRevB.51.6411}
}

@article{Li_2016_PRL_muSR_YbMgGaO4,
  title     = {Muon Spin Relaxation Evidence for the {U}(1) Quantum Spin-Liquid Ground State in the Triangular Antiferromagnet {YbMgGaO}$_4$},
  author    = {Li, Yuesheng and Adroja, Devashibhai and Biswas, Pabitra K. and Baker, Peter J. and Zhang, Qian and Liu, Juanjuan and Tsirlin, Alexander A. and Gegenwart, Philipp and Zhang, Qingming},
  journal   = {Physical Review Letters},
  volume    = {117},
  number    = {9},
  pages     = {097201},
  year      = {2016},
  doi       = {10.1103/PhysRevLett.117.097201}
}

@article{Paddison_2017_NatPhys,
  title   = {Continuous excitations of the triangular-lattice quantum spin liquid {YbMgGaO}$_4$},
  author  = {Paddison, Joseph A. M. and Daum, Marcus and Dun, Zhiling and Ehlers, Georg and Liu, Yaohua and Stone, Matthew B. and Zhou, Haidong and Mourigal, Martin},
  journal = {Nature Physics},
  volume  = {13},
  number  = {2},
  pages   = {117--122},
  year    = {2017},
  doi     = {10.1038/nphys3971}
}

@article{Shen_2016_Nature,
  title   = {Evidence for a spinon {F}ermi surface in a triangular-lattice quantum-spin-liquid candidate},
  author  = {Shen, Yao and Li, Yao-Dong and Wo, Hongliang and Li, Yuesheng and Shen, Shoudong and Pan, Bingying and Wang, Qisi and Walker, H. C. and Steffens, P. and Boehm, M. and Hao, Yiqing and Quintero-Castro, D. L. and Harriger, L. W. and Frontzek, M. D. and Hao, Lijie and Meng, Siqin and Zhang, Qingming and Chen, Gang and Zhao, Jun},
  journal = {Nature},
  volume  = {540},
  number  = {7634},
  pages   = {559--562},
  year    = {2016},
  doi     = {10.1038/nature20614}
}

@article{Li_2017_NatCommun,
  title     = {Nearest-neighbour resonating valence bonds in {YbMgGaO}$_4$},
  author    = {Li, Yuesheng and Adroja, Devashibhai and Voneshen, David and Bewley, Robert I. and Zhang, Qingming and Tsirlin, Alexander A. and Gegenwart, Philipp},
  journal   = {Nature Communications},
  volume    = {8},
  pages     = {15814},
  year      = {2017},
  doi       = {10.1038/ncomms15814}
}

@article{FisherWeichmanGrinsteinFisher1989,
  title = {Boson localization and the superfluid-insulator transition},
  author = {Fisher, Matthew P. A. and Weichman, Peter B. and Grinstein, G. and Fisher, Daniel S.},
  journal = {Phys. Rev. B},
  volume = {40},
  issue = {1},
  pages = {546--570},
  year = {1989},
  month = {Jul},
  publisher = {American Physical Society},
  doi = {10.1103/PhysRevB.40.546}
}

@article{Sondhi1997QPT,
  title = {Continuous quantum phase transitions},
  author = {Sondhi, S. L. and Girvin, S. M. and Carini, J. P. and Shahar, D.},
  journal = {Rev. Mod. Phys.},
  volume = {69},
  issue = {1},
  pages = {315--333},
  year = {1997},
  month = {Jan},
  publisher = {American Physical Society},
  doi = {10.1103/RevModPhys.69.315}
}

@article{Huckestein1995QHE,
  title = {Scaling theory of the integer quantum Hall effect},
  author = {Huckestein, Bodo},
  journal = {Rev. Mod. Phys.},
  volume = {67},
  issue = {2},
  pages = {357--396},
  year = {1995},
  month = {Apr},
  publisher = {American Physical Society},
  doi = {10.1103/RevModPhys.67.357}
}

@article{IgloiMonthus2005,
  title = {Strong disorder RG approach of random systems},
  author = {Igl{\'o}i, Ferenc and Monthus, C{\'e}cile},
  journal = {Physics Reports},
  volume = {412},
  number = {5--6},
  pages = {277--431},
  year = {2005},
  doi = {10.1016/j.physrep.2005.02.006}
}

\clearpage
%\appendix 
\onecolumngrid
\begin{center}
\textbf{\large Supplementary Material for\\
``Quantum criticality at strong randomness: a lesson from anomaly''}
\end{center}

\section{Anomaly and correlation in replica theory}
We follow the standard Euclidean path integral treatment of a disordered quantum system using replica trick. The object to consider is the partition function of $N$ identical copies of the system, all coupled with the disorder potential. The $N$-replicated partition function takes the form
\begin{equation}
\label{eq:replica}
    Z^{(N)}=\int Dv\prod_{a=1}^ND\phi_a \exp\left(-\sum_{a=1}^NS[\phi_a,v]-S_D[v] \right),
\end{equation}
where $\phi_a$ represents the dynamical degrees of freedom in each replica and $v$ is the random disorder potential. $S[\phi_a,v]$ describes the dynamics of each replica and its coupling to the disorder, while $S_D[v]$ describes the probability distribution of the disorder potential. The standard practice is to calculate correlation functions from the replicated partition function, with the replica limit $N\to0$ taken at the end of the calculation\cite{AltlandBook}.

The exact and average symmetries act differently in the replicated theory: the exact symmetry $K$ acts in a single replica, effectively becomes $K^{\otimes N}$, and the random potential $v$ transforms trivially under $K$; the average symmetry $G$, on the other hand, acts simultaneously in all replicas and $v$ can transform nontrivially under $G$. If the original symmetry is $K\times G$ (which is what we mainly focus on in this work), then in the replica theory the symmetry becomes $K^{\otimes N}\times G$. It is also useful to define the ``average part'' of the exact symmetry, which is the diagonal subgroup $K_{\rm diag}\in K^{\otimes N}$ -- this is the formal manifestation of the fact that an exact symmetry is automatically also an average symmetry. 

An anomaly between exact and average symmetry can now be understood as a mixed anomaly between $K^{\otimes N}$ and $G$ (see Refs.~\cite{MaWang2023,MZBCW2025} for additional subtleties). Moreover, for the anomalies considered in this work such as the average LSM, the anomaly becomes trivial if we restrict to the $K_{\rm diag}\times G$ subgroup -- namely if the exact symmetry becomes average.

Treating the replicated theory as an ordinary field theory,  we apply the ``power-law rule'' which demands some charged local operators to have nontrivial correlation functions. For the average symmetry $G$, we can simply consider a field $\phi^A_{a,\gamma}$ in any replica index $a$ that transforms nontrivially under $G$ with charge $\gamma$. The replica-symmetrized version is simply $\sum_a\phi^A_a$. The corresponding correlation function is 
\begin{align}
    \lim_{N \to 0}\langle \sum_{a}\phi^{A}_{a, \gamma} \sum_{b}\phi^{A}_{b, -\gamma} \rangle = \overline{\langle \phi^{A}_{\gamma}\phi^{A}_{-\gamma} \rangle},
\end{align}
which is just the ordinary first-moment correlation function.

Now for the exact symmetry $K$, an obvious candidate of charged operator is an operator $\phi^E_{a,\kappa}$ that is charged under $K^a$ in any replica index $a$ with charge $\kappa$. However, this operator is also charged under $K_{\rm diag}$, the average part of $K$. Since the anomaly is trivial when $K^{\otimes N}$ is restricted to $K_{\rm diag}$, this additional ``average charge'' of $\phi_a^E$ may cause short-range correlation. Instead, we should consider operators that are charged under $K^{\otimes N}$ but not $K_{\rm diag}$. The simplest choice is $\phi^E_{a,\kappa}\phi^E_{b,-\kappa}$ for $a\neq b$, or the replica-symmetrized version $\sum_{a\neq b}\phi^E_{a,\kappa}\phi^E_{b,-\kappa}$. The corresponding correlation function is
\begin{align}
    \lim_{N \to 0}\langle \sum_{a\neq b,a'\neq b'}\phi^{E}_{a, \kappa}\phi^{E}_{b, \kappa'} \phi^{E}_{a', -\kappa}\phi^{E}_{b', -\kappa'}\rangle = \overline{\left|\langle \phi^{E}_{\kappa}\phi^{E}_{-\kappa} \rangle\right|^2}.
\end{align}
which is the Edwards--Anderson correlator.

\section{Further analysis in the representative examples} 
\label{sec:more on examples}

\subsection{Anomaly constrained correlations}
\subsubsection{Strong disorder RG for random antiferromagnetic Heisenberg chain}
\label{sec:sdrg for spin chain}

To extend our results for the random antiferromagnetic Heisenberg chain to larger system sizes, we use techniques based on the strong-disorder renormalization group (SDRG). The SDRG approach encompasses several related schemes that we will discuss below. While the 1d and 2d Majorana models are also amenable to an SDRG analysis, these models are non-interacting, allowing us to explore sufficiently large system sizes through exact diagonalization. Below, we briefly describe the numerical SDRG protocol we used to study the random AFM spin-half Heisenberg chain. 

We begin with a lattice with even number of lattice sites, with a spin-half on each site. The starting Hamiltonian is $H = \sum_i J_{i,i+1} S_i \cdot S_{i+1}$. In the numerical implementation, we checked for both open and periodic boundary conditions, and found no qualitative difference in the correlations. The coupling on each bond is positive, $J_{i,i+1} > 0$. The SDRG framework proceeds as follows. Consider the strongest hopping element $J_{n, n+1} \equiv \Omega $. We treat $\Omega/J_{ij}$ as a large parameter in the Hamiltonian, and decimate the strongest bond, by projecting the decimated pair of spins to their singlet state. This decimation generates a new bond between spins on site $n-1$ and $n+1$, with an effective coupling that can be deduced from second order perturbation theory,
\begin{align}
    & H \to H + J_{n-1, n+2}^{\text{eff}} S_{n-1}\cdot S_{n+2} \\& J_{n-1, n+2}^{\text{eff}} = \frac{J_{n-1,n}J_{n+1, n+2}}{2\Omega} < J_{n-1,n}, J_{n+1,n+2} < \Omega.
\end{align}
This step has reduced the number of active bonds by 2 (decimates 3 bonds and generates 1, except for the last step). We can iterate this procedure until all the bonds are decimated, ultimately leading to a pattern of non-crossing random singlets. This procedure is well controlled under the assumption $\frac{J_{n,n-1}J_{n, n+1}}{2\Omega} \ll J_{n,n-1}, J_{n,n+1}$, and the fixed point limit is reached when $\frac{J_{n,n-1}J_{n, n+1}}{2\Omega} \to 0$, which is the infinite randomness fixed point, as rigorously established to be the stable RG fixed point by Fisher~\cite{PhysRevB.50.3799}. We implement this procedure numerically to extract the dimer-dimer correlations, averaged over samples.

\begin{figure}[h!]
    \centering
    \includegraphics[width=0.4\linewidth]{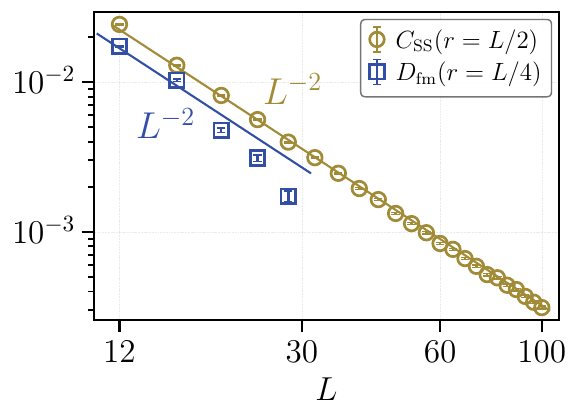}
    \caption{\justifying Power-law finite-size scaling of correlation functions in the random antiferromagnetic Heisenberg chain for operators charged under exact symmetries (spin-spin correlation $C_{\mathrm{SS}}$ [Eq.~\ref{eq:CSS}]) and average symmetries (first-moment dimer-dimer correlation $D_{\mathrm{fm}}$ [Eq.~\ref{eq:Dfm-def}]). The results are obtained using strong-disorder RG with $10^5$ disorder realizations, and we only show data with $r\ge 3$ and $\mathrm{SNR}\ge 10$.}
    \label{fig:SDRG-main-corr}
\end{figure}

Our numerical result for the random AFM Heisenberg chain is shown in Fig.~\ref{fig:SDRG-main-corr}. Using SDRG lets us access system sizes up to $L = 100$, while in the case of DMRG we stopped at $L=32$. We note, however, that the correlation $C_{\mathrm{SS}}(L)$ (with $r/L$ fixed) has the expected scaling of $L^{-2}$ from either method. Furthermore, the first-moment dimer-dimer correlation $D_{\mathrm{fm}}$ only has high SNR for sufficiently small values of $L$. Therefore, the range of $L$ values where we obtain reliable data for $D_{\mathrm{fm}}$ is almost the same for both DMRG and SDRG. In each case, we find the scaling $D_{\mathrm{fm}}(L) \sim L^{-2}$ (with fixed $r/L$).

%%%%%%%%%%%%%%%%%%%%%%%%%%%%%%%%%%%%%%
\subsubsection{Reliability of numerical evidence in $d=2$ and limitations of analytical SDRG}
\label{sec:2d numerics extra}
For our $1d$ examples, we complement numerics with an analytical strong-disorder renormalization group (SDRG) argument that predicts and explains the power-law exponents of the correlations of symmetry-charged operators discussed in this work (Sec.~\ref{appsec:sdrg_argument}). Developing a comparable analytical SDRG treatment in $2d$ is substantially more difficult. \cite{PhysRevB.108.064201}
In $2d$, real-space SDRG decimations generically generate effective couplings beyond nearest neighbors and quickly turn the original regular lattice into an irregular graph whose edges represent the effective interactions, including increasingly long-ranged links. This rapid proliferation of couplings, together with the fact that this interaction graph itself evolves under RG, makes it difficult to formulate an analytically tractable RG flow for the coupling statistics, since the decimation steps do not preserve a closed set of short-ranged interactions on a fixed lattice. It also complicates the arguments available in $1d$ at infinite-randomness fixed points that track which RG-generated configurations, and with what probability, contribute at leading order to a given two-point correlator. Consequently, SDRG studies of $2d$ disordered quantum criticality are typically pursued as numerical SDRG implementations that explicitly track the evolving graph and extract scaling from large-scale samples. \cite{Igloi2018,PhysRevB.82.054437, PhysRevX.8.041040}

\begin{figure*}[h!]
     \centering
     \begin{subfigure}[t]{0.5\textwidth}
        \centering
        \includegraphics[width=0.8\linewidth]{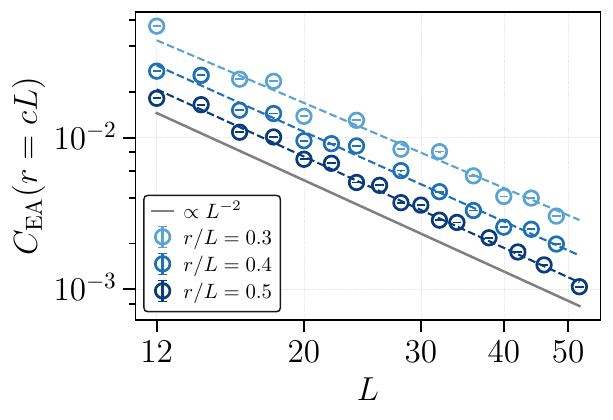}
        \caption{\centering Edwards--Anderson correlation}
    \end{subfigure}%
    %\hspace{2cm}
    \hfill
     \begin{subfigure}[t]{
     0.5\textwidth}
        \centering
        \includegraphics[width=0.8\linewidth]{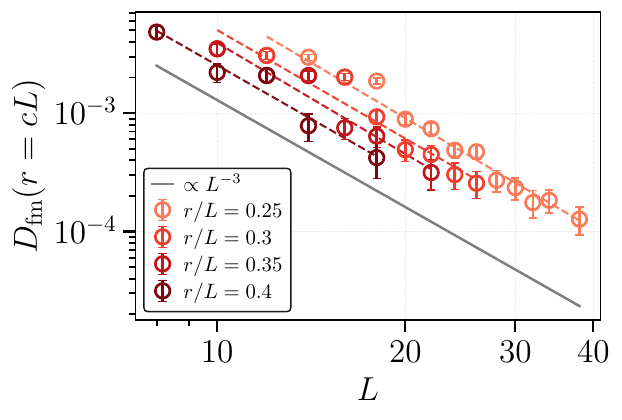}
        \caption{\centering First-moment dimer-dimer correlation}
    \end{subfigure}%
    \caption{\justifying Power-law finite-size scaling of correlation functions in the $2d$ random Majorana lattice for (a) the Edwards--Anderson correlator $C_{\mathrm{EA}}$ [Eq.~\ref{eq:EAcorrelator}],
    and (b) the first-moment dimer-dimer correlator $ D_{\mathrm{fm}}$ [Eq.~\ref{eq:Dfm-def}]. We obtain the data using exact diagonalization with $200L$ disorder realizations and show only points with $r\ge 3$, with the signal-to-noise ratio (SNR) above $2$ in (a) and above $3$ in (b).}
    \label{Fig:main-powerlaw-various-r-L}
\end{figure*}

Nevertheless, even without a comparably sharp analytical SDRG prediction in $2d$, our exact free-fermion numerics for the $2d$ random Majorana lattice provide a clear test of the power-law rule. In Fig.~\ref{Fig:main-powerlaw-various-r-L}, we perform finite-size scaling at fixed ratio $c=r/L$ for multiple values of $c$. We find robust power-law behavior across these different ratios; the Edwards--Anderson correlator follows
$C_{\mathrm{EA}}(r)\sim r^{-2}$, while the first-moment dimer-dimer correlator follows
$D_{\mathrm{fm}}(r)\sim r^{-3}$.
The consistency of the extracted exponents across several fixed $r/L$ values indicates that the observed scaling is not tied to a special choice of separation, and supports the reliability of our $2d$ numerical conclusions at the system sizes accessible to exact diagonalization.

Finally, we note that for the first-moment dimer-dimer correlation function, extracting the staggered ($k=\pi$) component from a finite set of disorder realizations is noticeably noisier in $d=2$, and the raw estimator $D_{\mathrm{fm}}(r)$ can exhibit a slowly varying offset.

To robustly isolate the $\pi$-momentum component in this case, we apply a finite-difference filter and define
\begin{equation}
\widetilde{D}_{\mathrm{fm}}(r)
=D_{\mathrm{fm}}(r)-
\frac{1}{2}[D_{\mathrm{fm}}(r-1)+D_{\mathrm{fm}}(r+1)].
\label{eq:D-fm-2d}
\end{equation}
which is the quantity used in our $d=2$ numerical analysis.
%%%%%%%%%%%%%%%%%%%%%%%%%%%%%%%%%%%%%%
\subsection{Other correlation functions of symmetry charged operators in the literature}
In this section we comment on other correlation functions studied in the literature, involving operators charged under both exact and average symmetries. We also discuss why these correlations are not constrained by anomalies.
%%%%%%%%%%%%%%%%%%%%%%%%%%%%%%%%%%%%%%
\subsubsection{Correlations of exact symmetry charged operators}

The simplest correlation function in the random Majorana hopping model is the disorder-averaged two-point function
\begin{equation}
    G(r)=\frac{1}{L^d}\sum_j \overline{\big\langle i\gamma_j\gamma_{j+r}\big\rangle}\, .
    \label{eq:def-G-r}
\end{equation}

The exact on-site symmetry is fermion parity, under which $\gamma_j$ is charged. Importantly, however, $G(r)$ is not an anomaly-enforced diagnostic. {Although for each sample realization, the correlator $\langle i\gamma_j\gamma_{j+r}\rangle$ may decay slowly (leading to slowly decaying EA correlator), the overall sign of the correlator may fluctuate strongly from sample to sample, leading to short-ranged first-moment correlator. Such behavior is quite generally anticipated for fermion systems with a finite mean-free path. } 

\begin{figure}[t]
  \centering
  \begin{subfigure}[t]{0.5\linewidth}
    \centering
    \caption{\centering Random Majorana lattice, $d=1$}
    \includegraphics[width=0.8\linewidth]{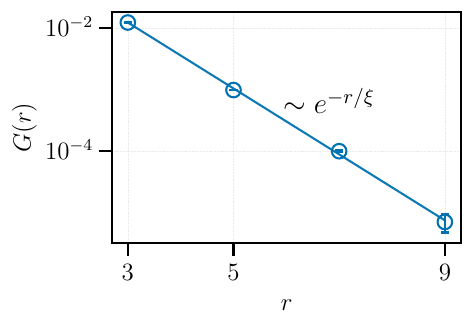}
  \end{subfigure}\hfill
  \begin{subfigure}[t]{0.5\linewidth}
    \centering
    \caption{\centering Random Majorana lattice, $d=2$}
    \includegraphics[width=0.8\linewidth]{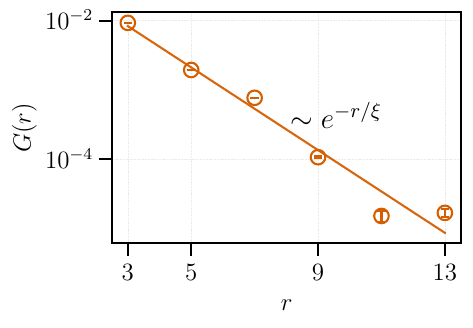}
  \end{subfigure}
  \caption{\justifying Exponential decay of the first-moment correlator $G(r)$ [Eq.~\ref{eq:def-G-r}] in the random Majorana lattice (semilog-$y$ plot): (a) $d=1$, $L=100$ with $10^{5}$ disorder realizations; (b) $d=2$, $L=52$ with $10^{4}$ disorder realizations. The results are obtained using exact diagonalization, and we show only data with $\mathrm{SNR}\ge 3$.}
  \label{fig:Gr-exp-decay-Majorana}
\end{figure}

We verify this expectation numerically in both $d=1$ and $d=2$ as shown in Fig.~\ref{fig:Gr-exp-decay-Majorana}. Despite the presence of an average LSM anomaly in these models, $G(r)$ exhibits a clear exponential decay, in agreement with our theoretical expectation.

%%%%%%%%%%%%%%%%%%%%%%%%%%%%%%%%%%%%%%
\subsubsection{Correlations of average symmetry charged operators}
Previous works have studied another dimer correlator that we refer to as the \emph{sample-connected dimer-dimer correlator} $D_{\mathrm{sc}}(r)$, in random-singlet phases realized in spin models in both $1d$~\cite{Shu_2016} and $2d$~\cite{PhysRevX.8.041040}. Let $d_j$ be the dimer operator defined in the main text, whose $\pi$-momentum component is charged under the average translation symmetry. The sample-connected correlator is defined by taking the connected correlator within each disorder realization and then averaging over realizations, 
\begin{equation}
    D_{\mathrm{sc}}(r)
    = \frac{1}{L^d}\sum_j \overline{
    \Bigl(
    \langle d_j d_{j+r}\rangle
    - \langle d_j\rangle \langle d_{j+r}\rangle
    \Bigr)}
    (-1)^r .
    \label{eq:D-sc-def}
\end{equation}
In Refs~\cite{Shu_2016,PhysRevX.8.041040}, $D_{\mathrm{sc}}(r)$ was shown to behave as $\sim 1/r^4$ in random singlet phases in both $1d$ and $2d$.

For the $1d$ random Majorana model, the sample-connected dimer correlator, after Jordan-Wigner transform, is related to the sample-connected correlator of the $\Z_2$-symmetric operator in random transverse-field Ising model. This correlator was studied in Ref.~\cite{Isingcorrelation}, and a similar $1/r^4$ behavior was found.

This correlator is not anomaly constrained. The anomaly-based power-law rule concerns correlators that, after disorder averaging, still retain the long-distance signature of the symmetry-breaking tendency associated with the anomalous symmetry. Here, $D_{\mathrm{sc}}(r)$ is defined to be \emph{connected} within each disorder realization, so the subtraction in Eq.~\eqref{eq:D-sc-def} removes the disconnected contribution that would otherwise carry the static long-distance imprint of dimer order in that realization. Consequently, $D_{\mathrm{sc}}(r)$ does not serve as an order diagnostic at long distances, and the anomaly does not impose a power-law form for it, so an exponential decay is allowed.

Nevertheless, $D_{\mathrm{sc}}(r)$ can still show a power-law decay in a strongly disordered critical state. Such a state can exhibit scale-free fluctuations even when the corresponding order parameter averages to zero and even when the anomaly does not constrain this observable. In this regime, $D_{\mathrm{sc}}(r)$ probes connected fluctuations of the staggered dimer density, and its long-distance behavior is controlled by the statistics of rare long-range structures generated by the strong-disorder physics rather than by symmetry-breaking order itself.

\begin{figure}[!htbp]
  \centering

  % ---------------- Top row: Majorana ----------------
  \begin{subfigure}[t]{0.5\linewidth}
    \centering
    \caption{\centering Random Majorana lattice, $d=1$}
    \includegraphics[width=0.8\linewidth]{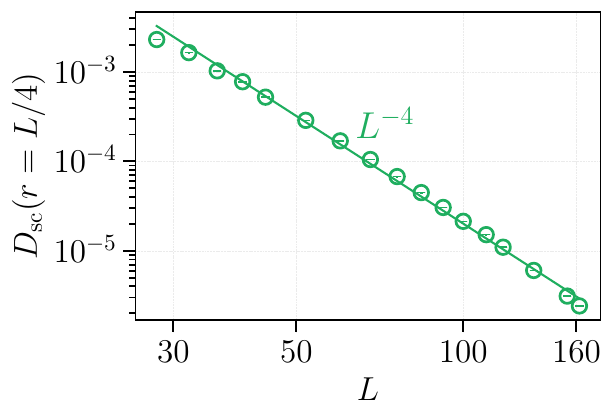}
  \end{subfigure}\hfill
  \begin{subfigure}[t]{0.5\linewidth}
    \centering
    \caption{\centering Random Majorana lattice, $d=2$}
    \includegraphics[width=0.8\linewidth]{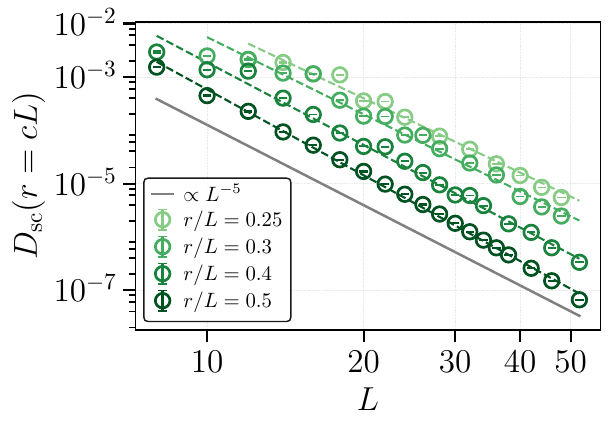}
  \end{subfigure}

  \vspace{0.6em}  % vertical separation between rows

  % ---------------- Bottom row: Heisenberg ----------------
  \begin{subfigure}[t]{0.5\linewidth}
    \centering
    \caption{\centering Random AFM Heisenberg chain, DMRG}
    \includegraphics[width=0.8\linewidth]{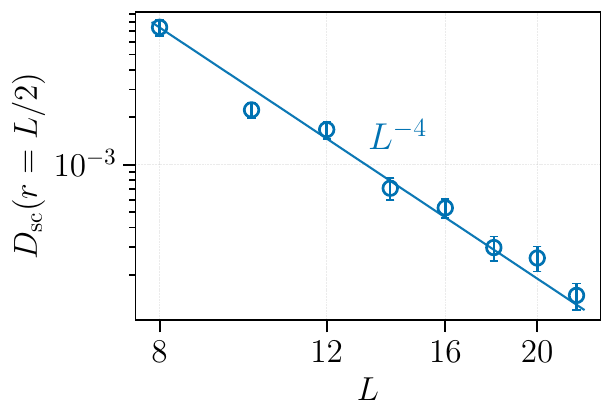}
  \end{subfigure}\hfill
  \begin{subfigure}[t]{0.5\linewidth}
    \centering
    \caption{\centering Random AFM Heisenberg chain, SDRG}
    \includegraphics[width=0.8\linewidth]{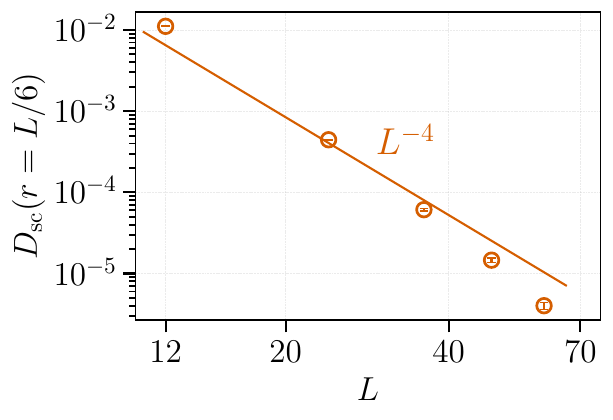}
  \end{subfigure}
  \caption{\justifying Finite-size scaling of the sample-connected dimer-dimer correlation function $D_{\mathrm{sc}}$ [Eq.~\ref{eq:D-sc-def}], showing power-law behavior. Top row: random Majorana lattice in (a) $d=1$ and (b) $d=2$. We obtain the data using exact diagonalization and show only points with $r \ge 3$ and $\mathrm{SNR}\ge 3$. Bottom row: random antiferromagnetic Heisenberg chain, with data obtained using two methods: (c) DMRG, showing only points with $r \ge 2$ and $\mathrm{SNR}\ge 5$, and (d) strong-disorder RG, showing only points with $r \ge 2$ and $\mathrm{SNR}\ge 10$.
}
  \label{fig:SC-combined}
\end{figure}

We observe this behavior numerically in the random antiferromagnetic Heisenberg chain and in the $d=1,2$ random Majorana lattice, as shown in Fig.~\ref{fig:SC-combined}. In the $1d$ cases, the data are well described by $D_{\mathrm{sc}}(r)\sim r^{-4}$, while in $2d$ the finite-size scaling at various fixed ratios $r/L$ is consistent with $D_{\mathrm{sc}}(r=cL)\sim L^{-5}$.
In $1d$, we present a theoretical understanding of the $r^{-4}$ scaling using an SDRG argument in Sec.~\ref{appsec:sdrg_argument}.

%%%%%%%%%%%%%%%%%%%%%%%%%%%%%%%%%%%%%%
\newpage
\section{Noise analysis and self-averaging of correlators}\label{appsec:snr_analysis}
A practically relevant question for disorder-averaged correlators is whether they are \emph{self-averaging}. 
For an observable $O(r,L)$ (for instance, a two-point correlator at separation $r$ in a system of size $L$), we estimate its disorder mean $\overline{O(r,L)}$ and its standard error $\mathrm{SE}[O(r,L)]$ from a finite ensemble of disorder realizations.  
We then use the standard signal-to-noise ratio defined as
\begin{equation}
    \mathrm{SNR}(r,L)\equiv \frac{\big|\overline{O(r,L)}\big|}{\mathrm{SE}\!\left[O(r,L)\right]},
\end{equation}

Self-averaging can be probed in two related but practically distinct ways. First, one may ask whether the disorder averaging at a fixed separation becomes sharp as the system grows: for fixed $r$,
\begin{equation}
    \frac{\mathrm{SE}[O(r,L)]}{\overline{O(r,L)}}\xrightarrow[L\to\infty]{}0,
\end{equation}
equivalently $\mathrm{SNR}(r,L)\to\infty$. 
Second, one may ask how the statistical precision at long distances behaves at fixed system size, i.e.\ how $\mathrm{SNR}(r,L)$ changes as $r$ increases at fixed $L$.
The latter does not define self-averaging in the strict sense, but it quantifies the range of $r$ over which a correlator can be reliably extracted at a given $L$ and sample count.
With this in mind, we examine $\mathrm{SNR}(r,L)$ in two complementary cuts of parameter space: (i) fixed $r$ while varying $L$, and (ii) fixed $L$ while varying $r$. 

\begin{figure}[!htbp]
    \centering
    % Panel (a)
    % a)
    \begin{subfigure}{\linewidth}
        \centering
        \subcaption{\justifying Spin-spin correlation, $C_{\mathrm{SS}}$ [Eq.~\ref{eq:CSS}]}
        \includegraphics[width=0.8\linewidth]{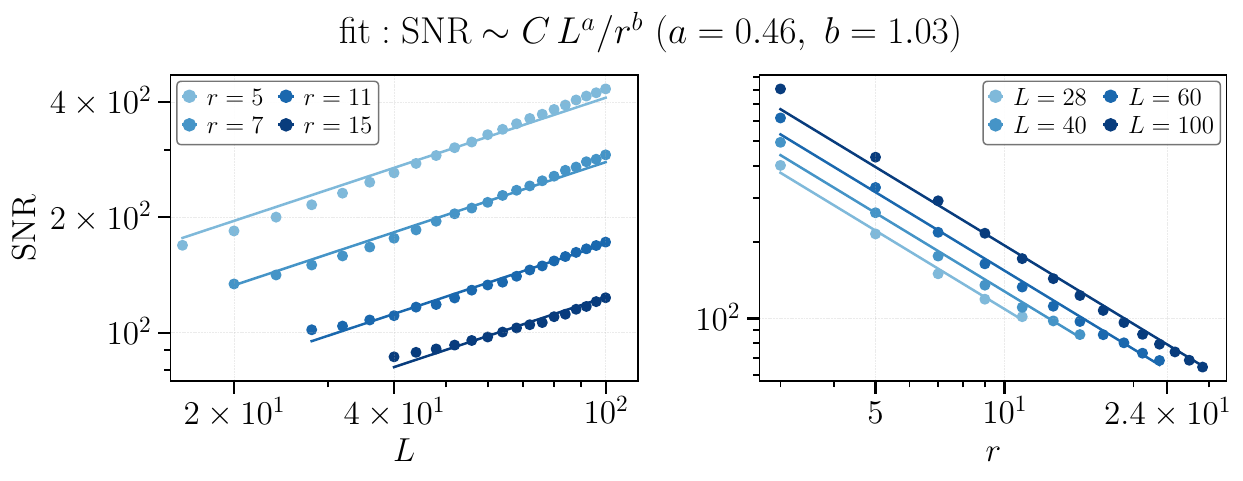}
        
        \label{fig:afm-snr_spin_spin}
    \end{subfigure}
    
    % Panel (b)
    \begin{subfigure}{\linewidth}
        \centering
        \subcaption{\justifying First-moment dimer-dimer correlation, $D_{\mathrm{fm}}$ [Eq.~\ref{eq:Dfm-def}]}
        \includegraphics[width=0.8\linewidth]{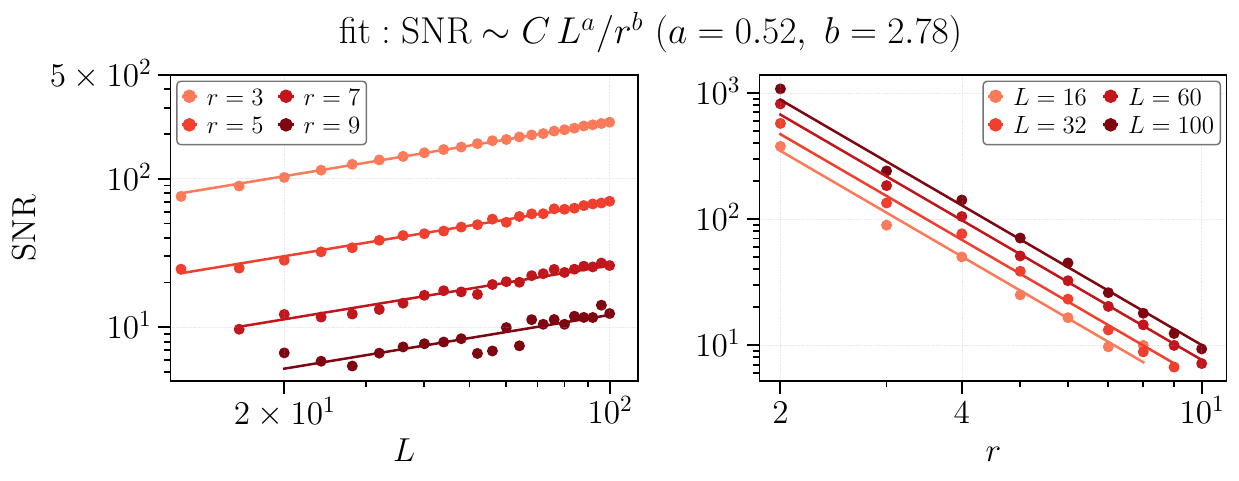}
        
        \label{fig:afm-snr_dimer_dimer_sample}
    \end{subfigure}

    % Panel (c)
    \begin{subfigure}{\linewidth}
        \centering
        \subcaption{\justifying Sample-connected dimer-dimer correlation, $D_{\mathrm{sc}}$ [Eq.~\ref{eq:D-sc-def}]}
        \includegraphics[width=0.8\linewidth]{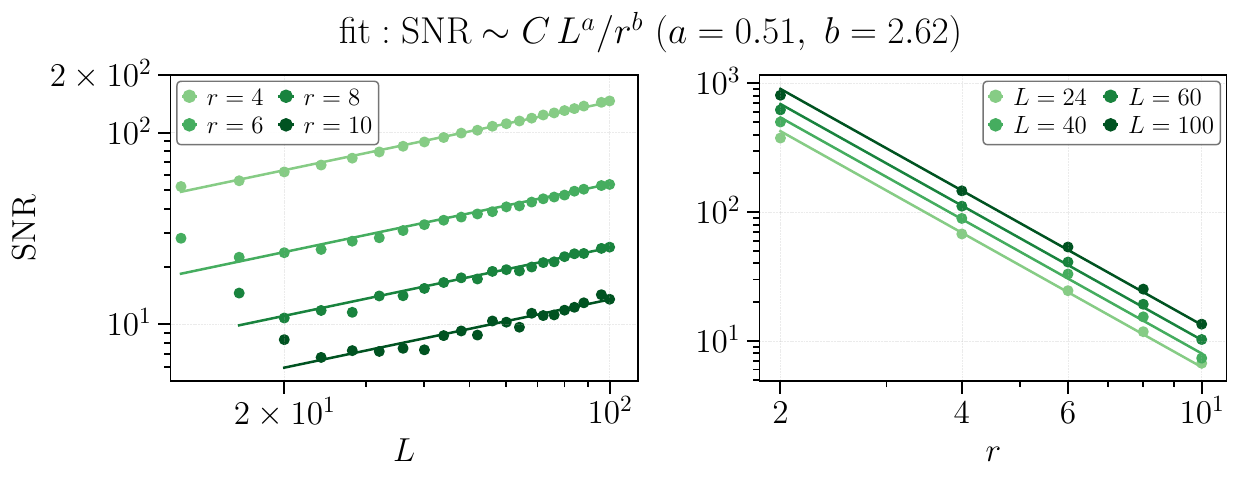}
        
        \label{fig:afm-snr_dimer_dimer_sample}
    \end{subfigure}
    
    \caption{\justifying Signal-to-noise ratio (SNR) analysis for power-law--decaying correlators in the random antiferromagnetic Heisenberg chain, computed using strong-disorder RG with $10^{5}$ disorder realizations. Left panels show $\mathrm{SNR}$ versus $L$ at fixed separation $r$; right panels show $\mathrm{SNR}$ versus $r$ at fixed system size $L$.
    The fit uses $20\le L\le 100$ and $3\le r\le 40$, with $\mathrm{SNR}\ge 5$.}
    \label{fig:SDRG_SNR}
\end{figure}

\begin{figure}[!htbp]
    \centering
    
    % Panel (a)
    % a)
    \begin{subfigure}{\linewidth}
        \centering
        \subcaption{\justifying Edwards--Anderson correlation, $C_{\mathrm{EA}}$ [Eq.~\ref{eq:EAcorrelator}]}
        \includegraphics[width=0.8\linewidth]{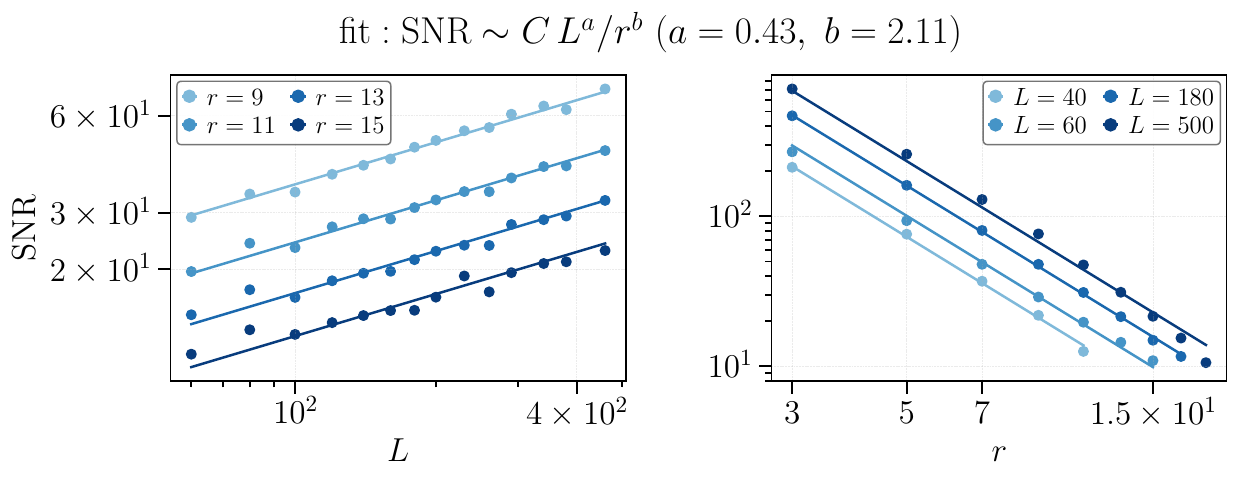}
        
        \label{fig:1}
    \end{subfigure}
    
    % Panel (b)
    \begin{subfigure}{\linewidth}
        \centering
        \subcaption{\justifying First-moment dimer-dimer correlation, $D_{\mathrm{fm}}$ [Eq.~\ref{eq:Dfm-def}]}
        \includegraphics[width=0.8\linewidth]{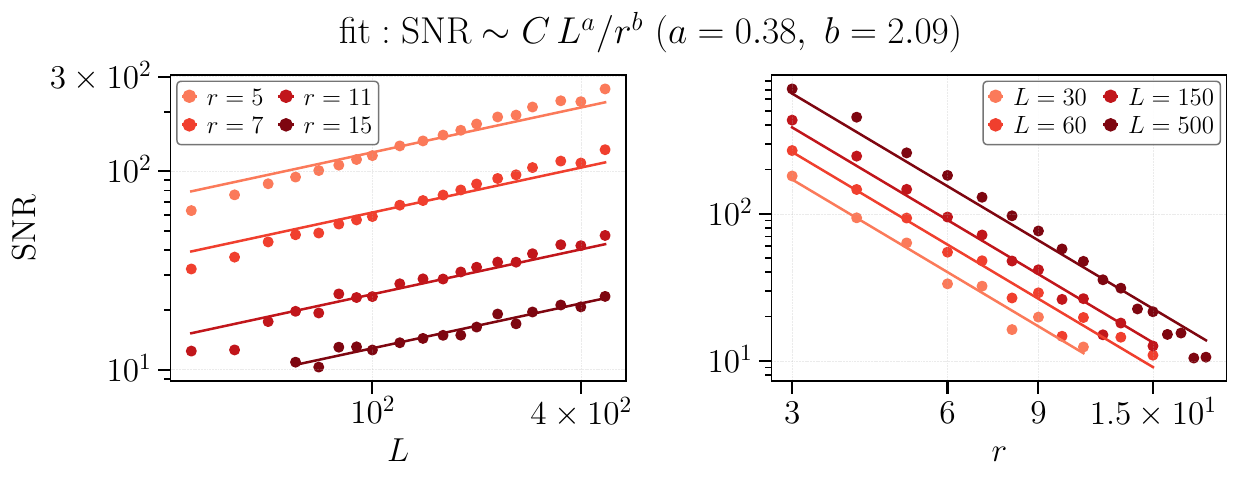}
        
        \label{fig:2}
    \end{subfigure}

    % Panel (c)
    \begin{subfigure}{\linewidth}
        \centering
        \subcaption{\justifying Sample-connected dimer-dimer correlation, $D_{\mathrm{sc}}$ [Eq.~\ref{eq:D-sc-def}]}
        \includegraphics[width=0.8\linewidth]{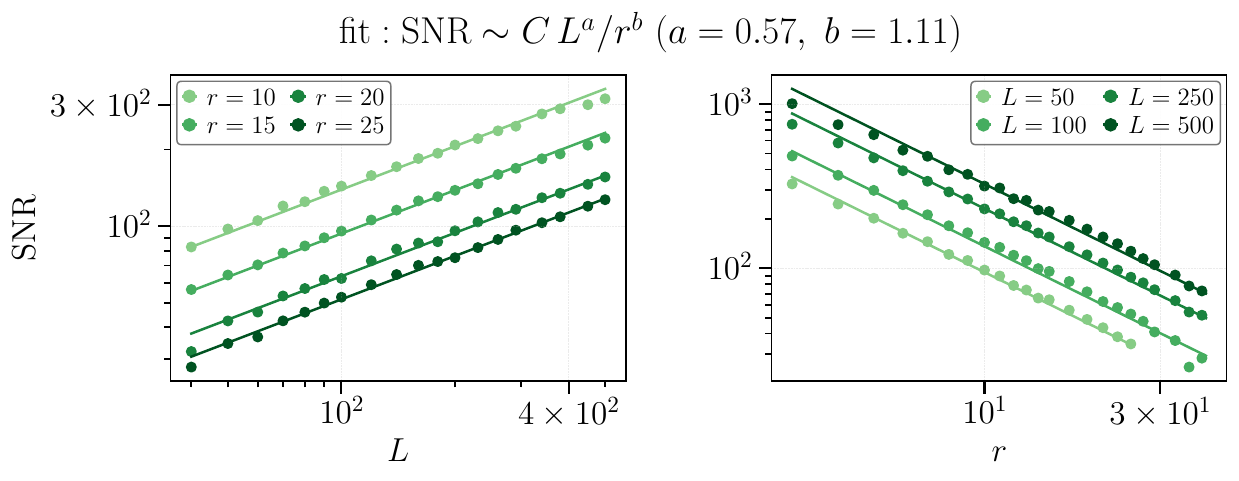}
        
        \label{fig:3}
    \end{subfigure}
    \caption{\justifying Signal-to-noise ratio (SNR) analysis for power-law--decaying correlators in the $1d$ random Majorana lattice, computed using exact diagonalization with $5000$ disorder realizations. Left panels show $\mathrm{SNR}$ versus $L$ at fixed separation $r$; right panels show $\mathrm{SNR}$ versus $r$ at fixed system size $L$. The fit uses $40 \le L \le 500$ and $3 \le r \le 40$, with $\mathrm{SNR} \ge 10$.}
    \label{fig:1dMajorana_SNR}
\end{figure}

\begin{figure}[!htbp]
    \centering
    
    % Panel (a)
    % a)
    \begin{subfigure}{\linewidth}
        \centering
        \subcaption{\justifying Edwards--Anderson correlation, $C_{\mathrm{EA}}$ [Eq.~\ref{eq:EAcorrelator}]}
        \includegraphics[width=0.8\linewidth]{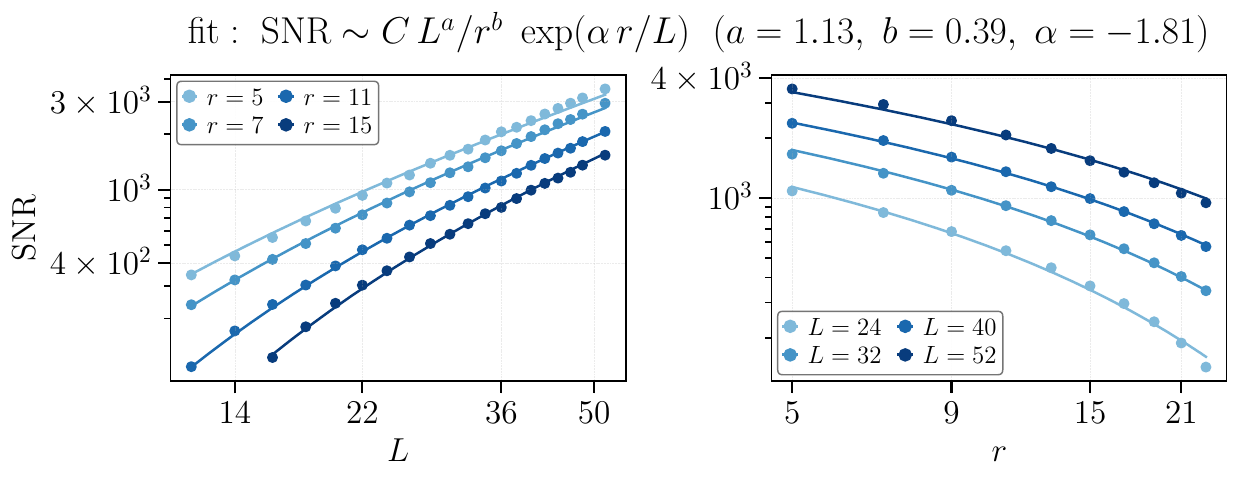}
        
        \label{fig:1}
    \end{subfigure}
    
    % Panel (b)
    \begin{subfigure}{\linewidth}
        \centering
        \subcaption{\justifying First-moment dimer-dimer correlation, ${D}_{\mathrm{fm}}$ [Eq.~\ref{eq:D-fm-2d}]}
        \includegraphics[width=0.8\linewidth]{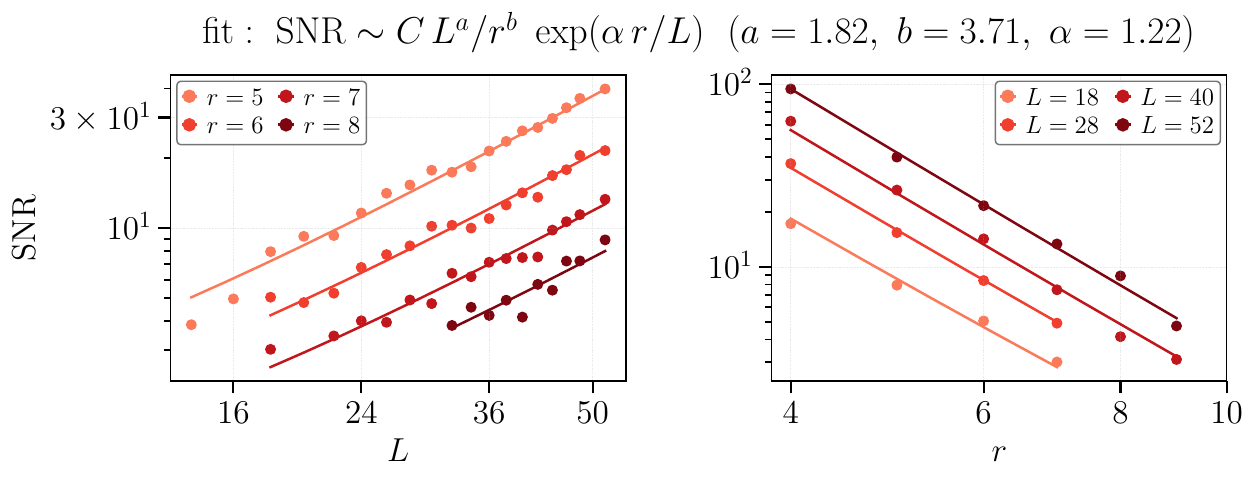}
        
        \label{fig:2}
    \end{subfigure}

    % Panel (c)
    \begin{subfigure}{\linewidth}
        \centering
        \subcaption{\justifying Sample-connected dimer-dimer correlations, $D_{\mathrm{sc}}$ [Eq.~\ref{eq:D-sc-def}]}
        \includegraphics[width=0.8\linewidth]{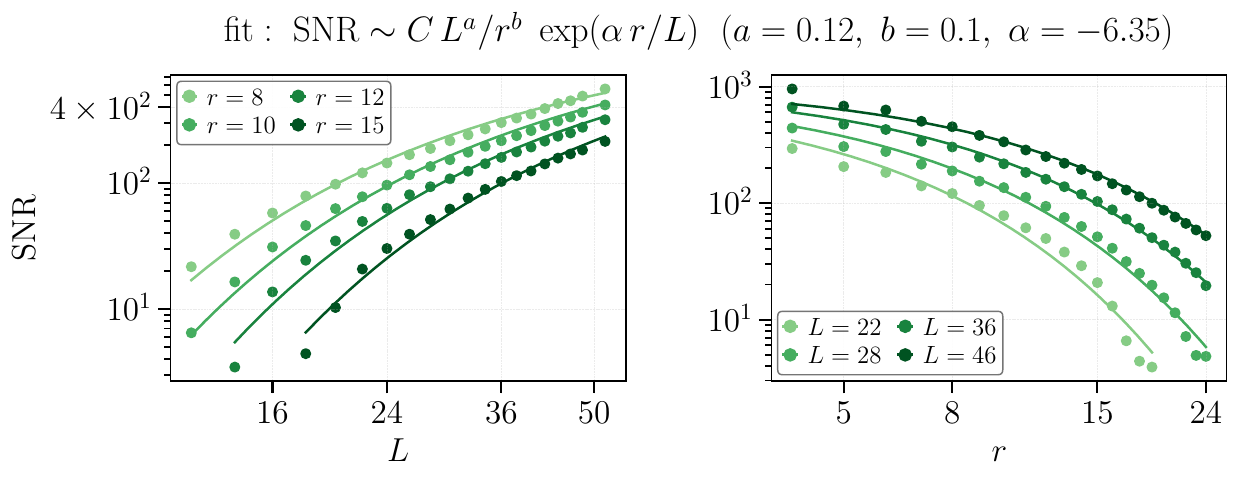}
        
        \label{fig:3}
    \end{subfigure}
    \caption{\justifying Signal-to-noise ratio (SNR) analysis for power-law--decaying correlators in the $2d$ random Majorana lattice, computed using exact diagonalization with $4000$ disorder realizations. Left panels show $\mathrm{SNR}$ versus $L$ at fixed separation $r$; right panels show $\mathrm{SNR}$ versus $r$ at fixed system size $L$. The fit uses $12 \le L \le 52$ and $4 \le r \le 24$, with $\mathrm{SNR} \ge 3$.}
    \label{fig:2dMajorana_SNR}
\end{figure}

For the $1d$ models, the results are shown in Figs.~\ref{fig:SDRG_SNR} and \ref{fig:1dMajorana_SNR} for the random antiferromagnetic Heisenberg chain and the $1d$ random Majorana lattice, respectively. 
For every anomaly-constrained correlator of a symmetry-charged operator that we analyze, the data are consistent with an asymptotic form
\begin{equation}\label{eq:snr_1d_scaling}
    \mathrm{SNR}(r,L)\sim \frac{L^{a}}{r^{b}},\qquad b>a>0,
\end{equation}
within the fitting windows stated in the figure captions.
The fitted exponent $a>0$ indicates self-averaging in the fixed-separation sense, since $\mathrm{SNR}(r,L)$ increases with $L$ when $r$ is held fixed, while $b>0$ implies that at fixed $L$ the SNR decreases with distance and large separations are statistically harder to resolve. 
The additional inequality $b>a$ sharpens this statement. Along separations that scale with system size, $r=\kappa L$, one has
\begin{equation}
    \mathrm{SNR}(\kappa L,L)\sim \kappa^{-b}L^{a-b}\to 0 \qquad (L\to\infty),
\end{equation}
so the loss of SNR with distance outpaces the gain from increasing $L$. 
Equivalently, to maintain a fixed target SNR while probing separations that grow with $L$, one would need to keep $r/L$ sufficiently small or increase the number of disorder realizations accordingly, rather than relying on larger $L$ alone. 
Since $\mathrm{SE}\propto N_{\mathrm{samp}}^{-1/2}$, one has $\mathrm{SNR}\propto \sqrt{N_{\mathrm{samp}}}$ at fixed $(r,L)$, so maintaining $\mathrm{SNR}(\kappa L,L)$ at a fixed target value requires $N_{\mathrm{samp}}\propto L^{2(b-a)}$ up to model-dependent prefactors.
We find similar behavior for other power-law--decaying correlators of symmetry-charged operators that are not anomaly-constrained, such as the sample-connected dimer-dimer correlator also shown in Figs.~\ref{fig:SDRG_SNR} and \ref{fig:1dMajorana_SNR}.
\FloatBarrier
% \FloatBarrier
\vspace*{-0.1\baselineskip}
For the $2d$ random Majorana lattice, Eq.~\eqref{eq:snr_1d_scaling} provides a reasonable baseline fit, but the residuals retain a systematic dependence on the aspect ratio $r/L$.
This indicates that, at the system sizes accessible here, the approach to the asymptotic regime depends not only on $r$ and $L$ separately, but also on the finite-size ratio $r/L$.
Rather than imposing an explicit $r/L$ cutoff, which would introduce a direct bias towards a specific power-law scaling regime, we incorporate the leading finite-size correction as a multiplicative scaling function $f(r/L)$ and fit
\begin{equation}\label{eq:snr_2d_finite_size}
    \mathrm{SNR}(r,L)\sim \frac{L^{a}}{r^{b}}\,f(r/L),
    \qquad 
    f(r/L)\approx \exp\!\big(\alpha\, r/L\big).
\end{equation}
This modification captures the curvature seen in $\log\mathrm{SNR}$ versus $\log r$ at fixed $L$ while preserving the expected large-$L$ limit of the $d=1$ ansatz.
At fixed $r$ and $L\to\infty$, the ratio $r/L\to 0$ so $f(r/L)\to 1$, and the scaling reduces to a pure power-law. Figure~\ref{fig:2dMajorana_SNR} shows that Eq.~\eqref{eq:snr_2d_finite_size} provides a consistent description for the anomaly-constrained Edwards--Anderson and first-moment dimer-dimer correlators in the $2d$ Majorana lattice model, and it also fits the sample-connected dimer-dimer correlator, with fit parameters listed in the figure. 

We note that in our $2d$ results, the first-moment dimer-dimer correlator exhibits $b>a$ in Fig.~\ref{fig:2dMajorana_SNR}(b), so along separations that scale with system size, such as $r=\kappa L$, one expects $\mathrm{SNR}(\kappa L,L)\sim \kappa^{-b}L^{a-b}$ and hence a decreasing SNR with increasing $L$ along $r\propto L$. 
By contrast, the Edwards--Anderson correlator in Fig.~\ref{fig:2dMajorana_SNR}(a) yields $b<a$ within the shown window, while the sample-connected dimer-dimer correlator in Fig.~\ref{fig:2dMajorana_SNR}(c) gives exponents closer to $b\simeq a$. 
These differences emphasize that $b>a$ is not a uniform feature of all $d=2$ observables at the sizes accessible here. These behaviors of the SNR motivates keeping $L_{\max}\approx 50$ in the $d=2$ analysis, since pushing to substantially larger $L$ while maintaining comparable statistical control at large separations would require a much larger number of disorder realizations.

The sign of $\alpha$ is informative about the leading finite-size drift with $c=r/L$.
For the Edwards--Anderson correlator and the sample-connected dimer-dimer correlator we find $\alpha<0$ [Fig.~\ref{fig:2dMajorana_SNR}(a,c)], indicating an additional suppression of the SNR as one probes separations that are a larger fraction of the system size, beyond what is captured by the baseline $L^{a}/r^{b}$ scaling.
This effect is especially pronounced for $D_{\mathrm{sc}}$, where the large negative $\alpha$ shows that the SNR deteriorates rapidly as $r/L$ increases.
For the first-moment dimer-dimer correlator we find $\alpha>0$ [Fig.~\ref{fig:2dMajorana_SNR}(b)], so the $r/L$-dependent correction bends in the opposite direction and partially compensates that loss over the fitting window.

Overall, the SNR analysis serves as a quantitative sanity check on the regime in which disorder-averaged correlators can be reliably interpreted as exhibiting power-law scaling, given the system sizes and disorder ensembles accessible in our numerics.

\FloatBarrier

%%%%%%%%%%%%
\newpage
\section{Strong disorder RG argument for the dimer-dimer correlation functions}
\label{appsec:sdrg_argument}
In this section we present an SDRG argument that supports our numerical results for the dimer-dimer correlations in the 1$d$ random-singlet phase of the Majorana or spin models we study. 

As discussed in the main text, we construct dimer operators that are charged (i.e. their expectation values transform by a sign) under the average translation symmetry.

It is natural to parametrize the dimer operator as
\begin{equation}
d_i \;\approx\;  (-1)^i\,\varepsilon_i ,
\end{equation}
where the expectation value $\overline{\langle d_i\rangle}$ vanishes, and $\varepsilon_i$ is translation-invariant. This makes it clear that any long-distance connected dimer-dimer correlation function has the asymptotic form
\begin{equation}
\bigl\langle d_id_{i+r}\bigr\rangle_{\rm conn}
\;\sim\; (-1)^r F(r),
\end{equation}
with a real function $F(r)$ that (in the examples we study) decays with $r$.  In what follows we use SDRG to determine this power-law decay for systems of random-singlet type.

In the SDRG procedure, at each step we identify the currently strongest bond and denote its coupling by $\Omega$. We decimate this bond by forming a singlet and removing the two spins from the low-energy Hilbert space. As the RG proceeds, more spins are eliminated and the remaining active sites become increasingly sparse. If the strongest bond at the $n$th decimation step is $\Omega_n$, we parametrize the RG flow by the scale $\Gamma = -\frac{\log \Omega}{\log \Omega_n}$.

If we start at the random-singlet fixed point, the arguments of Fisher show that the
density of surviving (active) spins at RG scale $\Gamma$ behaves
as $n(\Gamma)\sim \Gamma^{-2}$.\cite{PhysRevB.50.3799}. So if the chain initially has $N_0$
sites, the number of active spins is $N(\Gamma)\sim N_0/\Gamma^{2}$.
The probability that a site is active at scale $\Gamma$ is
therefore
\begin{equation}
p_{\text{surv}}(\Gamma) \;=\; \frac{N(\Gamma)}{N_0}
\;\sim\; \Gamma^{-2},
\label{eq:psurv}
\end{equation}
The same fixed-point solution implies that the typical spacing between
active spins grows as $\ell(\Gamma)\sim \Gamma^{2}$.  To probe
correlations at separation $r$ we should thus consider the RG scale
$\Gamma_r$ where $\ell(\Gamma_r)\sim r$, i.e.
\begin{equation}
\Gamma_r^{2}\sim r \quad\Rightarrow\quad \Gamma_r\sim \sqrt{r},
\end{equation}
Inserting this into Eq.~\eqref{eq:psurv} gives
\begin{equation}
p_{\text{surv}}(\Gamma_r)\sim \Gamma_r^{-2}\sim r^{-1},
\end{equation}
so each site has probability of order $1/r$ to remain active up to the
stage at which singlets of length $\sim r$ are being formed.

\noindent From this perspective there are three qualitatively distinct ways the
SDRG flow can treat the two dimers at $(i,i+1)$ and $(i+r,i+r+1)$:\\
(a) The most common configuration is that all four spins in
$(i,i+1)$ and $(i+r,i+r+1)$ are decimated separately at early RG scales, forming short singlets with nearby neighbours; by the time singlets of length $\sim r$ appear, there is no active degree of freedom left in either
dimer region, and the two bonds are effectively independent. Since
each site survives up to scale $\Gamma_r$ only with probability
$p_{\text{surv}}(\Gamma_r)\sim 1/r$, the probability that any of the
four spins is still active at that scale is of order $1/r$.  Thus, the
probability that all four have already been decimated is
$ P_{(a)}(r) = 1 - \mathcal O\!\left(\frac{1}{r}\right),$
so this configuration occurs with probability of order 1,
$P_{(a)}(r)\approx 1$, for large $r$.\\
(b) More rarely, one effective
spin from the vicinity of $(i,i+1)$ and one from $(i+r,i+r+1)$ both survive until they become nearest neighbours among the active sites and form a single long singlet of length $\sim r$; this requires two
specific sites to survive and therefore occurs with probability
$P_{(b)}(r)\sim p_{\text{surv}}(\Gamma_r)^2\sim r^{-2}$. \\
(c) Even more rarely, all four spins associated with the two dimers (or their RG descendants) survive to scale $\Gamma_r$ and form two distinct long singlets spanning the two dimer regions; this requires four sites to survive and occurs with probability
$P_{(c)}(r)\sim p_{\text{surv}}(\Gamma_r)^4\sim r^{-4}$.

We first consider the disorder-averaged dimer-dimer correlation
$\overline{\langle d_i d_{i+r}\rangle}$.  In the SDRG picture this average can be expressed as a weighted sum over the three
configurations (a)--(c),
\begin{equation}
\overline{\langle d_i d_{i+r}\rangle}
= \sum_{k=a,b,c} P_{(k)}(r)\,\mathcal{D}_{(k)},
\end{equation}
where $P_{(k)}(r)$ is the probability of case $(k)$ and 
$\mathcal{D}_{(k)}$ is the corresponding (order--one) value of 
$\langle d_i d_{i+r}\rangle$ in that configuration.
In the typical configuration (a) both dimers are short-range bonds, so
$\langle d_i d_{i+r}\rangle$ is controlled by local physics in the two
dimer regions and is independent of $r$ once it is large.  The rarer
configurations (b) and (c) shift this local value only by an amount of
order one, but their contributions are suppressed by their small
probabilities, $P_{(b)}(r)\sim r^{-2}$ and $P_{(c)}(r)\sim r^{-4}$.
As a result,
\begin{equation}
\overline{\langle d_i d_{i+r}\rangle}
= \text{const.} + \mathcal O\!\left(\frac{1}{r^{2}}\right),
\end{equation}
and the unconnected disorder average is dominated by an
$r$-independent background; the staggered, $(-1)^r$--oscillating piece only appears in the subleading corrections.

The behavior changes once we consider the first-moment dimer-dimer correlator $D_{\mathrm{fm}}(r)$ defined in Eq.~\eqref{eq:Dfm-def}. For the SDRG estimate it is convenient to focus on a representative pair of dimers at separation $r$, for which Eq.~\eqref{eq:Dfm-def} reduces to
\begin{equation}
D_{\rm fm}(r)
= \left(\overline{\langle d_i d_{i+r}\rangle}
  - \overline{\langle d_i\rangle}\,\overline{\langle d_{i+r}\rangle}\right)(-1)^r,
\end{equation}
By translation invariance the disorder-averaged one-point functions
are site independent,
$\overline{\langle d_i\rangle}\approx\overline{\langle d_{i+r}\rangle}\equiv m$,
so
\begin{equation}
D_{\mathrm{fm}}(r)
= \sum_{k=a,b,c} P_{(k)}(r)\,\mathcal{D}_{(k)} - m^2
= \sum_{k=a,b,c} P_{(k)}(r)\,\Delta_{(k)},
\end{equation}
with $\Delta_{(k)} \equiv \mathcal{D}_{(k)} - m^2$.  Thus $D_{\text{fm}}(r)$ is
still a first-moment quantity over the three SDRG configurations, but with
shifted weights $\Delta_{(k)}$.  For the typical case (a) the two
dimer regions are effectively independent, so $\mathcal{D}_{(a)}\approx m^2$ and
hence $\Delta_{(a)}\approx 0$; meaning that the large $r$-independent contribution
that dominated $\overline{\langle d_i d_{i+r}\rangle}$ is removed by
the subtraction.  The remaining contribution is governed by the rare
cases (b) and (c), for which $\Delta_{(b)},\Delta_{(c)}=\mathcal
O(1)$ while $P_{(b)}(r)\sim r^{-2}$ and $P_{(c)}(r)\sim r^{-4}$.  In
case (b) the long singlet that connects the two dimer regions couples
to the staggered piece $(-1)^i\varepsilon_i$, so the surviving
contribution is staggered and decays with the probability
$P_{(b)}(r)$.  Combining this with the general structure discussed
above, we obtain
\begin{equation}
D_{\mathrm{fm}}(r)\sim \frac{A}{r^{2}}
 + \mathcal O\!\left(\frac{1}{r^{4}}\right),
\end{equation}

Finally, we consider the sample-connected dimer-dimer correlator $D_{\mathrm{sc}}(r)$ defined in Eq.~\eqref{eq:D-sc-def}. Again focusing on a representative pair of dimers at separation $r$, Eq.~\eqref{eq:D-sc-def} can be written as
\begin{equation}
    D_{\mathrm{sc}}(r)
    = \overline{
    \Bigl(
    \langle d_j d_{j+r}\rangle
    - \langle d_j\rangle \langle d_{j+r}\rangle
    \Bigr)}
    (-1)^r,
\end{equation}
In case (a) the two dimer regions are independent within each
realization, so
$\langle d_i d_{i+r}\rangle \approx \langle d_i\rangle
\langle d_{i+r}\rangle$ and the sample-connected correlation is
essentially zero.  In case (b) a single long singlet connects the two
regions; both $d_i$ and $d_{i+r}$ are controlled by the same two-spin
state, and to leading order
$\langle d_i d_{i+r}\rangle$ again factorizes into
$\langle d_i\rangle\langle d_{i+r}\rangle$, so this case also does
not contribute at leading order.  The minimal configuration that
produces a nonzero sample-connected contribution is  case
(c), with two distinct long singlets of length $\sim r$ spanning the
two dimers.  In this situation $d_i d_{i+r}$ contains cross-terms in
which one factor comes from singlet \#1 and the other from singlet
\#2; these are genuine four-spin contributions that cannot be written
as $\langle d_i\rangle\langle d_{i+r}\rangle$, and hence survive
in the sample-connected combination with an $\mathcal{O}(1)$ value.
Since such configurations occur with probability
$P_{(c)}(r)\sim r^{-4}$ and inherit the momentum-$\pi$ charge of the
dimer operator, the sample-connected correlation function decays as
\begin{equation}
D_{\text{sc}}(r)\sim \frac{B}{r^{4}}.
\end{equation}
in agreement with the numerical results.

\end{document}